\PassOptionsToPackage{table}{xcolor}

\documentclass[sigconf]{acmart}

\AtBeginDocument{%
  \providecommand\BibTeX{{%
    \normalfont B\kern-0.5em{\scshape i\kern-0.25em b}\kern-0.8em\TeX}}}

\setcopyright{acmcopyright}
\copyrightyear{2019} 
\acmYear{2019} 
\acmConference[CIKM '19]{The 28th ACM International Conference on Information and Knowledge Management}{November 3--7, 2019}{Beijing, China}
\acmBooktitle{The 28th ACM International Conference on Information and Knowledge Management (CIKM '19), November 3--7, 2019, Beijing, China}
\acmPrice{15.00}
\acmDOI{10.1145/3357384.3357882}
\acmISBN{978-1-4503-6976-3/19/11}
\usepackage{booktabs}
\usepackage{amsmath,amssymb,amsfonts}
\usepackage{algorithmic}
\usepackage{graphicx}
\usepackage{textcomp}
\usepackage{enumerate}
\usepackage{booktabs}
\usepackage{calc}
\usepackage{subfigure}
\usepackage{wrapfig}
\usepackage[keeplastbox]{flushend}
\usepackage[table]{xcolor}
\usepackage{colortbl}
\usepackage{framed}
\usepackage[framemethod=TikZ]{mdframed}
\usepackage{multirow}
\usepackage{mathrsfs}
\usepackage[ruled,vlined]{algorithm2e}
\usepackage{appendix}
\usepackage{balance}
\usepackage{bm}
\usepackage[misc]{ifsym}
\usepackage{bbding}

\newlength\maxWidth
\newtheorem{property}{\textbf{Property}}
\newtheorem{qcProof}{\textbf{Proof}}
\newtheorem{definition}{\textbf{Definition}}
\newtheorem{example}{\textbf{Example}}
\SetKwProg{Function}{Function}{}{}
\SetKwProg{Procedure}{Procedure}{}{}

\definecolor{boxColor}{RGB}{255,255,255}
\definecolor{grayColor}{gray}{0.95}
\definecolor{redColor}{RGB}{223,83,83}
\definecolor{greenColor}{RGB}{60,160,60}
\definecolor{whiteColor}{RGB}{255,255,255}

\mdfdefinestyle{myboxstyle}{
  linewidth=0.6pt,
  backgroundcolor=boxColor,
  leftline=true, rightline=false,
  topline=false, bottomline=false,
  innerleftmargin=2, innerrightmargin=2,
  innertopmargin=2, innerbottommargin=2,
}

\begin{document}
\fancyhead{}

\title{
  BePT: A Behavior-based Process Translator for Interpreting and Understanding Process Models
}

\author{Chen Qian}
\affiliation{
  \institution{Tsinghua University}
  \city{Beijing}
  \country{China}
}
\email{qc16@mails.tsinghua.com}

\author{Lijie Wen}
\authornote{Corresponding Author.}
\affiliation{
  \institution{Tsinghua University}
  \city{Beijing}
  \country{China}
}
\email{wenlj@tsinghua.edu.cn}

\author{Akhil Kumar}
\affiliation{
  \institution{Penn State University}
  \city{Pennsylvania}
  \country{USA}
}
\email{akhilkumar@psu.edu}

\begin{abstract}
  Sharing process models on the web has emerged as a common practice. Users can collect and share their experimental process models with others. However, some users always feel confused about the shared process models for lack of necessary guidelines or instructions. Therefore, several process translators have been proposed to explain the semantics of process models in natural language (NL). We find that previous studies suffer from information loss and generate semantically erroneous descriptions that diverge from original model behaviors. In this paper, we propose a novel process translator named \emph{BePT} (\textbf{Be}havior-based \textbf{P}rocess \textbf{T}ranslator) based on the encoder-decoder paradigm, encoding a process model into a middle representation and decoding the representation into NL descriptions. Our theoretical analysis demonstrates that \emph{BePT} satisfies behavior correctness, behavior completeness and description minimality. The qualitative and quantitative experiments show that \emph{BePT} outperforms the state-of-the-art baselines.
\end{abstract}




\maketitle

\section{Introduction} \label{sec:introduction}
\begin{figure}[t]
  \centering
  \includegraphics[width=0.95\maxWidth]{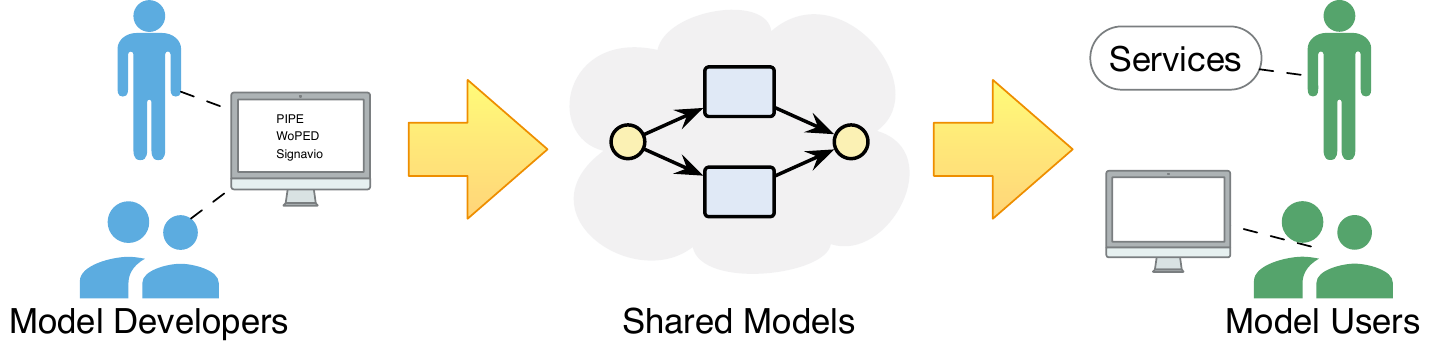}
  \caption{The Model Sharing Scenario: The users collect the shared models developed by the third-party developers.}
  \label{fig:senario}
  \end{figure}

A process consists of a series of interrelated tasks. Its graphical description is called \textbf{process model} \cite{processMining}. Over the past decade, a specific kind of process model - scientific workflow - has been established as a valuable means for scientists to create reproducible experiments \cite{ScientificWorkflows}. Several scientific workflow management systems (SWFM) have become freely available, easing scientific models' creation, management and execution \cite{ScientificWorkflows}. However, creating scientific models using SWFMs is still a laborious task and complex enough to impede non-computer-savvy researchers from using these tools \cite{future}. Therefore, cloud repositories emerged to allow sharing of process models, thus facilitating their reuse and repurpose \cite{myExperiment1, myExperiment2, ScientificWorkflows}. As shown in Figure \ref{fig:senario}, the model developers use modeling tools to build and manage process models which can be hosted in the cloud, then run as a service or be downloaded to users' local workspaces. Popular examples of such scientific model platforms or repositories include myExperiment, Galaxy, Kepler, CrowdLabs, Taverna, VisTrails, e-BioFlow, e-Science and SHIWA \cite{ScientificWorkflows, Taverna, Galaxy, myExperiment1, myExperiment2}. As for users, reusing shared models from public repositories is much more cost-effective than creating, testing and tuning a new one.

However, those models are difficult to reuse since they lack necessary NL guidelines or instructions to explain the steps, jump conditions and related resources \cite{requirement_text,Leo,Hen,Goun}. For example, the repository offered with myExperiment currently contains more than 3918 process models from various disciplines including bioinformatics, astrophysics, earth sciences and particle physics \cite{ScientificWorkflows}, but only 1293 out of them have corresponding NL documents\footnote{The data is collected from \url{https://www.myexperiment.org/} before Aug. 2019}, which shows the gap between the shared models and their NL descriptions. This real-world scenario illustrates that the cloud platforms do not have effective means to address this translation problem \cite{Goun,SharingModels}, i.e., automatically translating the semantics of process models into NL, thus making it challenging for users to reuse the shared models. Now that means to translate a process model are becoming available to help users understand models and improve shared models' reusability \cite{requirement_text}, a growing interest in exploring automatic process translators - \textbf{process to text (P2T)} techniques - has emerged.

We define our problem as follows: given a process model, our approach aims to generate the textual descriptions for the semantics of the model. We choose Petri nets as our modeling language for their: 1) formal semantics; 2) many analysis tools; 3) ease of transformation from/to other modeling languages \cite{petri_three_reason}. Our approach - \emph{BePT} - first embeds the structural and linguistic information of a model into a tree representation, and then linearizes it by extracting its behavior paths. Finally, it generates sentences for each path. Our theoretical analysis and the experiments we conducted demonstrate that \emph{BePT} satisfies three desirable properties and outperforms the state-of-the-art P2T approaches.

To summarize, our contributions are listed as follows:

\begin{enumerate}[1)]
  \item We propose a behavior-based process translator \emph{BePT} to generate textual descriptions without behavioral errors. To the best of our knowledge, this work is the first attempt that fully considers model behaviors in process translation.
  \item The "encoder-decoder" paradigm and unfolded behavior graphs are employed to help better analyze and extract correct behavior paths to be described in natural language.
  \item We formally analyze BePT's properties and conducted experiments on ten-time larger (compared with previous works) datasets collected from industry and academic fields to better reveal the statistical characteristics. The results demonstrate BePT's strong expressiveness and reproducibility.
\end{enumerate}

\section{Related Work}
\textbf{Path-based Process Translators}. The path-based approach \cite{Leo} was proposed to generate the text of a process model. It first extracts language information before annotating each label \cite{label1, label2, label3}. Then it generates the annotated tree structure before traversing it by \textit{depth first search}. Once sentence generation is triggered, it employs NL tools to generate corresponding NL sentences \cite{realpro}. This work solved the annotation problem, but it only works for structured models and ignores unstructured parts.

\textbf{Structure-based Process Translators}. Structure-based translator \cite{Hen} was subsequently proposed to handle unstructured parts. It recursively extracts the longest path of a model on the unstructured parts to linearize each activity. However, it only works on certain patterns and is hard to extend more complex situations. Along this line, another structure-based method was proposed \cite{Goun} which can handle more elements and complex patterns. It first preprocesses a model by trivially reversing loop edges and splitting multiple-entry-multiple-exit gateways. Then, it employs heuristic rules to match every source gateway node with the goal nodes. Next, it unfolds the original model based on those matched goals. Finally, it generates the texts of the unfolded models. Although this structure-based method maintains good paragraph indentations, it neglects the behavior correctness and completeness.

\textbf{Other Translators}. Other "to-text" works that take BPMN \cite{nlg_bpmn}, EPC \cite{nlg_epc}, UML \cite{nlg_uml}, image \cite{nlg_image} or video \cite{nlg_video} as inputs are difficult to apply into the process-related scenarios or are not for translation. Hence, we aim to design a novel process translator.

\section{Preliminaries} \label{sec:preliminaries}
Before going further into the main idea, we introduce some background knowledge: Petri net \cite{petri1, petri2, unstructuredModel}, \textbf{R}efined \textbf{P}rocess \textbf{S}tructure \textbf{T}ree (RPST) \cite{rpst}, \textbf{C}omplete \textbf{F}inite \textbf{P}refix (CFP) \cite{unfolding1, unfolding2, unfolding3} and \textbf{D}eep \textbf{Syn}tactic \textbf{T}ree (DSynT) \cite{realpro, dsyn}. These four concepts are respectively used for process modeling, structure analysis, behavior unfolding and sentence generation.

\begin{figure*}[hbtp]
  \centering
  \subfigure[A bioinformatics Petri net model ($N_1$).]{
    \label{sfig:pre_petri}
    \includegraphics[width=0.65\maxWidth]{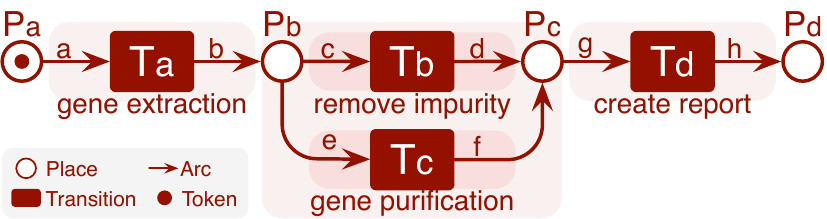}
  }
  \hspace{0.1cm}
  \subfigure[The RPST of $N_1$.]{
    \label{sfig:pre_rpst}
    \includegraphics[width=0.30\maxWidth]{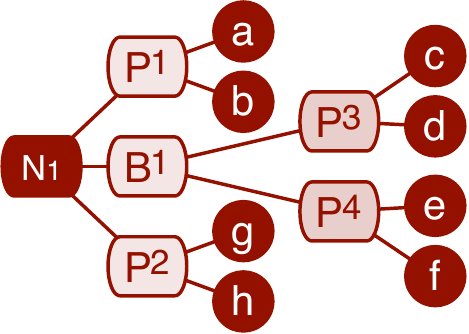}
  }
  \hspace{0.1cm}
  \subfigure[The CFP of $N_1$ ($\mathbb{N}_1$).]{
    \label{sfig:pre_cfp}
    \includegraphics[width=0.50\maxWidth]{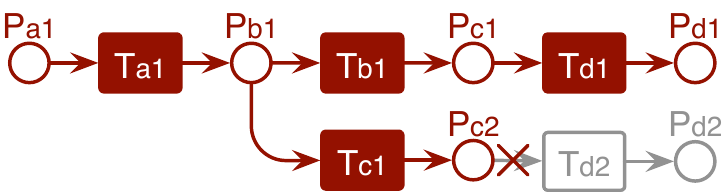}
  }
  \hspace{0.1cm}
  \subfigure[The DSynT of $T_a$ of $N_1$.]{
    \label{sfig:pre_dsynt}
    \includegraphics[width=0.30\maxWidth]{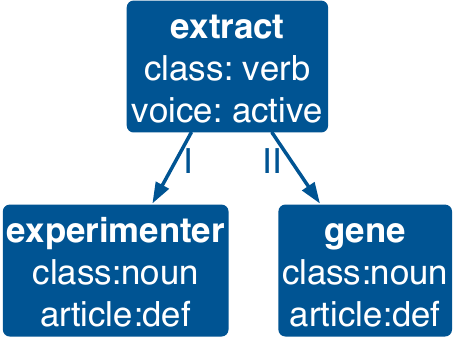}
  }
  \caption{An example of a bioinformatic process model.}
  \label{fig:preliminary}
\end{figure*}

\subsection{Petri Net}
\begin{definition}[Petri Net, Net System, Boundary node]
  A \textbf{Petri net} $N$ is a tuple $(P,T,F)$, where $P$ is a finite set of places, $T$ is a finite set of transitions, $F\subseteq (P\times T)\cup (T\times P)$ is a set of directed arcs. A marking of $N$, denoted $M$, is a bag of tokens over $P$. A \textbf{net system} $S=(N,M)$ is a Petri net $N$ with an initial marking $M$. The input set and output set of a node $n$ are respectively denoted as $\bullet n=\{x|(x,n)\in F\}$ and $n\bullet=\{x|(n,x)\in F\}$. The source and sink sets of a net $N$ are respectively denoted as $\bullet N=\{x \in P \cup T | \bullet x=\varnothing\}$ and $N\bullet=\{x \in P \cup T | x\bullet=\varnothing\}$. These boundary elements $\bullet N\bullet=\bullet N\cup N\bullet$ are called \textbf{boundary nodes} of $N$.
  \end{definition}
  
\begin{definition}[Firing Sequence, TAR, Trace]
  Let $S=(N,M)$ be a net system with $N=(P,T,F)$. A transition $t\in T$ can be \textbf{fired} under a marking $M$, denoted $(N,M)[t\rangle$, iff each $p\in \bullet t$ contains at least one token. After $t$ fires, the marking $M$ changes to $M \backslash \bullet t \cup t\bullet$ (Firing Rule). A sequence of transitions $\sigma=t_1 t_2\cdots t_{n}\in T^*$ is called a \textbf{firing sequence} iff $(N,M)[t_1\rangle(N,M_1)[t_2\rangle\cdots[t_n\rangle(N,M_n)$ holds. Any transition pair that fires contiguously ($t_i\prec t_{i+1}$) is called a \textbf{transition adjacency relation (TAR)}. A firing sequence $\sigma$ is a \textbf{trace} of $S$ iff the tokens completely flow from all source(s) to sink(s).
  \end{definition}
  
\begin{example}
  Figure \ref{sfig:pre_petri} shows a real-life bioinformatics process model expressed by Petri net. $P_a$ contains one token so that the current marking $M$ is $[1,0,0,0]$ (over $[P_a,P_b,P_c,P_d]$). According to the firing rule, each node in the input set of $T_a$ ($\bullet T_a=\{P_a\}$) contains at least one token so that $T_a$ can be fired. After firing $T_a$, the marking becomes $M \backslash \bullet T_a \cup T_a\bullet$, i.e., $[0,1,0,0]$. The TAR set of $N_1$ is $\{T_a\prec T_b, T_a\prec T_c, T_b\prec T_d, T_c\prec T_d\}$. The trace set of $N_1$ is $\{T_aT_bT_d, T_aT_cT_d\}$.
  \end{example}

\subsection{Refined Process Structure Tree (RPST)}
\begin{definition}[Component, RPST, Structured, Unstructured]
  A process \textbf{component} is a sub-graph of a process model with a single entry and a single exit (SESE), and it does not overlap with any other component. The \textbf{RPST} of a process model is the set of all the process components. Let $C=\bigcup_{i=1}^{n}\{c_i\}$ be a set of components of a process model. $C$ is a \textbf{trivial} component iff $C$ only contains a single arc; $C$ is a \textbf{polygon} component iff the exit node of $c_i$ is the entry node of $c_{i+1}$; $C$ is a \textbf{bond} component iff all sub-components share same boundary nodes; Otherwise, $C$ is a \textbf{rigid} component. A rigid component is a region of a process model that captures arbitrary structure. Hence, if a model contains no rigid components, we say it is \textbf{structured}, otherwise it is \textbf{unstructured}.
  \end{definition}

\begin{example}
  The colored backgrounds in Figure \ref{sfig:pre_petri} demonstrate the decomposed components which naturally form a tree structure - RPST - shown in Figure \ref{sfig:pre_rpst}. The whole net (polygon) can be decomposed into three first-layer SESE components ($P^1$, $B^1$, $P^2$), and these three components can be decomposed into second-layer components ($a$, $b$, $P^3$, $P^4$, $g$, $h$). The recursive decomposition ends at a single arc (trivial).
  \end{example}

\subsection{Complete Finite Prefix (CFP)}
\begin{definition}[Cut-off transition, mutual, CFP] \label{def:cfp}
  A branching process $O=(P,T,F)$ is a completely fired graph of a Petri net satisfying that 1) $|\bullet p|\le 1, \forall p\in P$; 2) no element is in conflict with itself; 3) for each $x$, the set $\{y\in P\cup T|y \prec x\}$ is finite. The \textbf{mapping function $\hbar$} maps each CFP element to the corresponding element in the original net. If two nodes $p_1, p_2$ in CFP satisfy $\hbar(p_1)=\hbar(p2)$, we say they are \textbf{mutual} (places) to each other.
  
  A transition $t$ is a \textbf{cut-off transition} if there exists another transition $t'$ such that $Cut([t])=Cut([t'])$ where $[t]$ denotes a set of transitions of $t$ satisfying TAR closure ($\forall e \in T : e \prec t \Rightarrow e \in [t]$) and $Cut([t])=\hbar(\bullet O\cup [t]\bullet \backslash \bullet [t])$. A \textbf{CFP} is the greatest backward closed subnet of a branching process containing no transitions after any cut-off transition.
  \end{definition}
  
\begin{example}
  Figure \ref{sfig:pre_cfp} shows the branching process of $N_1$ (including the light-gray part). Since each original node corresponds to one or more CFP nodes, thus we append "id" to number each CFP node. As $\hbar(P_{c1})=\hbar(P_{c2})=P_c$ so that $P_{c1}$ and $P_{c2}$ are mutual (places). In $\mathbb{N}_1$, $Cut([T_{b1}])=Cut([T_{c1}])=P_c$ so that $T_{c1}$ is a cut-off transition (transitions after $T_{c1}$ are cut). The cut graph is CFP of $N_1$ (excluding the light-gray part).
  \end{example}

\subsection{Deep Syntactic Tree (DSynT)}
  A DSynT is a dependency representation of a sentence. In a DSynT, each node carries a verb or noun decorated with meta information such as the tense of the main verb or the number of nouns etc, and each edge can denote three dependencies - subject (I), object (II), modifier (ATTR) - between two adjacency nodes.

\begin{example}
  Figure \ref{sfig:pre_dsynt} shows the DSynT of $T_a$ in $N_1$. The main verb ``\emph{extract}'' is decorated by class ``\emph{verb}'' and the voice ``\emph{active}''. The subject and the object of ``\emph{extract}'' are ``\emph{experimenter}'' (assigned by the model developer) and ``\emph{gene}''. This DSynT represents the dependency relations of the sentence ``\emph{the experimenter extracts the genes}''.
  \end{example}

\section{Our Method} \label{sec:method}
First, we list some non-trivial challenges to be solved:

\begin{enumerate}[\textbf{C}1]
  \item How to analyze and decompose the structure of a complex model, such as an unstructured or multi-layered one?
  \item For each model element, how to analyze the language pattern of a short label, extract the main linguistic information and create semantically correct descriptions?
  \item How to transform a non-linear process model into linear representations, especially when it contains complex patterns?
  \item How to extract the correct behaviors of process models and avoid behavior space explosion?
  \item How to design language templates and simplify textual descriptions to express more naturally? How to make the results more intuitive to read and understand?
  \item How to avoid semantic errors and redundant descriptions?
  \end{enumerate}

To solve these challenges (\textbf{C1}-\textbf{C6}), we propose \emph{BePT} which is built on the encoder-decoder framework inspired from machine translation systems \cite{machineTranslation1, machineTranslation2, machineTranslation3}. The encoder creates an intermediate tree representation from the original model and the decoder generates its NL descriptions. Figure \ref{fig:framework} presents a high-level framework of BePT, including four main phases: Structure Embedding, Language Embedding, Text Planning and Sentence Planning \cite{Leo, Hen, Goun}:

\begin{figure}[t]
  \centering
  \includegraphics[width=0.95\maxWidth]{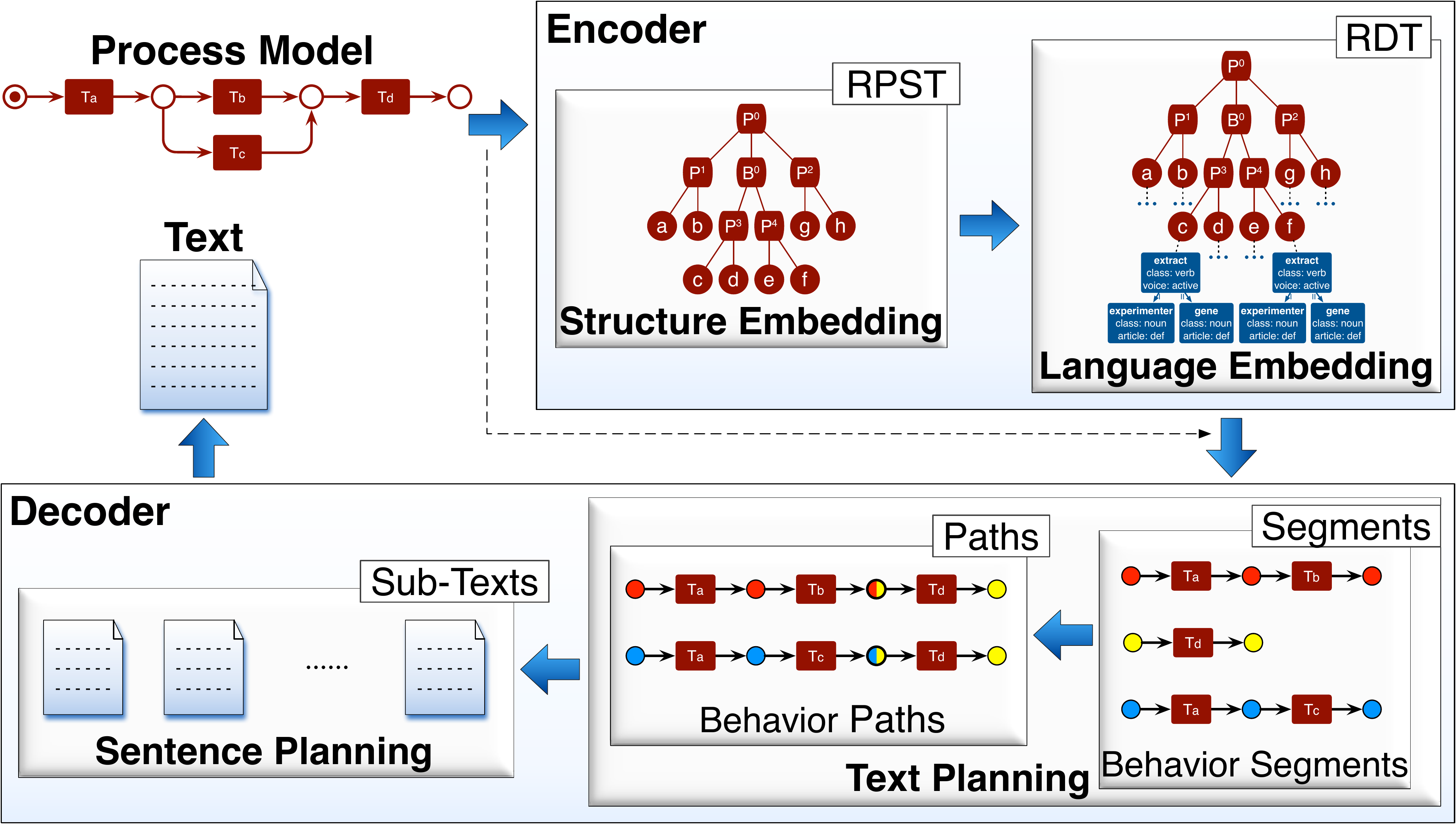}
  \caption{High-level view of BePT's framework.}
  \label{fig:framework}
  \end{figure}

\begin{enumerate}[1)]
  \item \textbf{\emph{Structure Embedding} (C1)}: Embedding the structure information of the original model into the intermediate representation.
  \item \textbf{\emph{Language Embedding} (C2)}: Embedding the language information of the original model into the intermediate representation.
  \item \textbf{\emph{Text Planning} (C3, C4)}: Linearizing the non-linear tree representation into linear representations.
  \item \textbf{\emph{Sentence Planning} (C5, C6)}: Generating NL text by employing pre-defined language templates and NL tools.
\end{enumerate}

\subsection{Structure Embedding}
\begin{figure}[hbtp]
  \includegraphics[width=0.70\maxWidth]{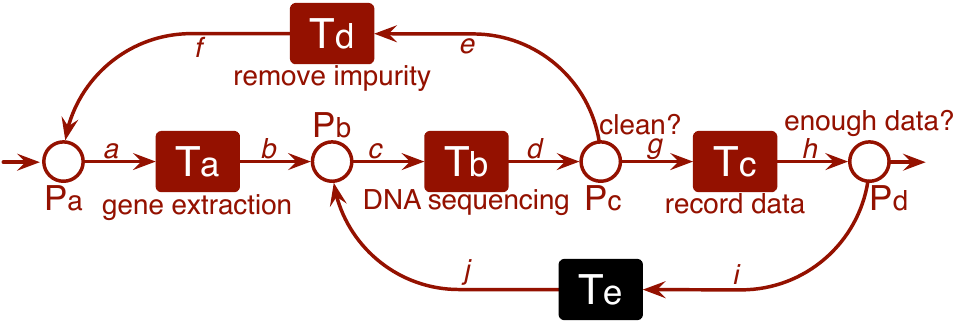}
  \caption{A simplified model ($N_*$). The original complex component (any structure is possible) is simplified by the black element (a single activity).}
  \label{fig:rigid}
  \end{figure}

We take a simplified model $N_*$, shown in Figure \ref{fig:rigid}, as our running example due to its complexity and representativeness. A complex sub-component (any structure is possible) in the original model is replaced by the black single activity $T_e$. The simplified model $N_*$ is also complex since it contains a main path and two loops.

We employ a simplification algorithm from \cite{Goun} to replace each sub-model with a single activity to obtain a simplified but behavior-equivalent one because a model containing many sub-models may complicate the behavior extraction \cite{Goun}. In the meantime, the simplification operation causes no information loss \cite{Goun} since the simplified part will be visited in the deeper recursion. We emphasize that this simplification step is easy and extremely necessary for behavior correctness (see Appendix \ref{prf:correctness} for proof of behavior correctness).

Next, we analyze its structural skeleton and then create the RPST of $N_*$. Finally, we embed its structural information - RPST - into a tree representation (as shown in the upper part of Figure \ref{fig:rdt}).

\subsection{Language Embedding}
\subsubsection{\textbf{Extract Linguistic Information}}
This step sets out to recognize NL labels and extract the main linguistic information \cite{label1, label2, label3}. For each NL label, we first examine prepositions and conjunctions. If prepositions or conjunctions are found, respective flags are set to true. Then we check if the label starts with a gerund. If the first word of the label has an ``\emph{ing}'' suffix, it is verified as a gerund verb phrase (e.g., ``\emph{extracting gene}''). Next, \emph{WordNet} \cite{wordnet} is used to learn if the first word is a verb. If so, the algorithm refers it to a verb phrase style (e.g., ``\emph{extract gene}''). In the opposite case, the algorithm proceeds to check prepositions in the label. A label containing prepositions the first of which is ``\emph{of}'' is qualified as a noun phrase with \emph{of} prepositional phrase (e.g., ``\emph{creation of database}''). If the label is categorized to none of the enumerated styles, the algorithm refers it to a noun phrase style (e.g., ``\emph{gene extraction}''). Finally, we similarly categorize each activity label into four labeling styles (gerund verb phrase, verb phrase, noun phrase, noun phrase with \emph{of} prepositional phrase).

Lastly, we extract the linguistic information - role, action and objects - depending on which pattern it triggers. For example, in $N_*$, the label of $T_d$ triggers a verb phrase style. Accordingly, the action lemma ``\emph{remove}'' and the noun lemma ``\emph{impurity}'' are extracted.

\subsubsection{\textbf{Create DSynTs}}
\begin{figure}[hbtp]
  \centering
  \includegraphics[width=0.65\maxWidth]{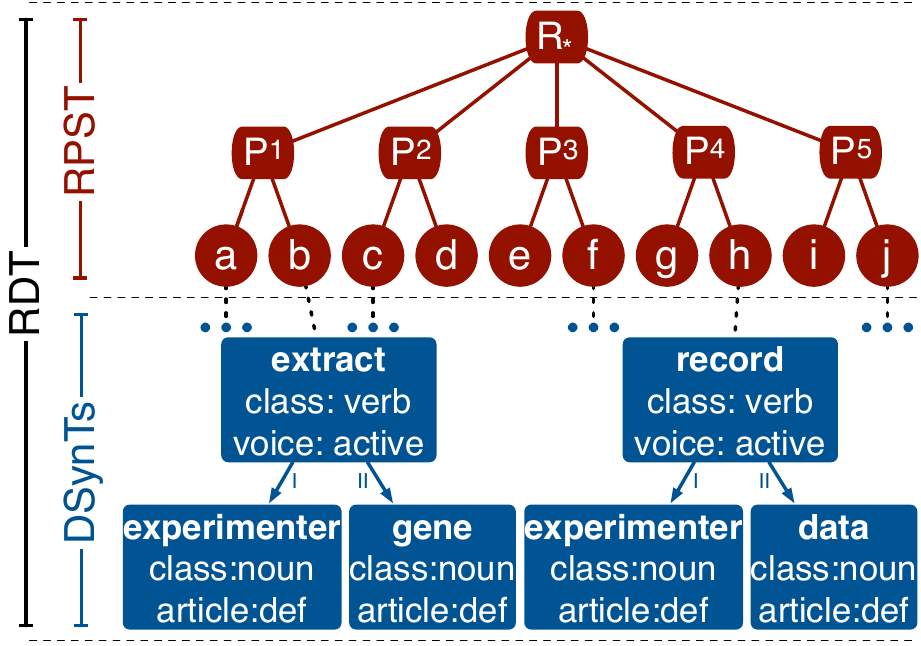}
  \caption{The RDT of $N_*$. Some parts are replaced by the ellipsis due to the limited space.}
  \label{fig:rdt}
\end{figure}

Once this main linguistic information is extracted, we create a DSynT for each label by assigning the main verb and main nouns including other associated meta information \cite{dsyn, realpro} (as shown in the lower part of Figure \ref{fig:rdt}).

For better representation, we concatenate each DSynT root node to its corresponding RPST leaf node, and we call this concatenated tree RPST-DSynTs (\textbf{RDT}). The RDT of $N_*$ is shown in Figure \ref{fig:rdt}.

So thus far, we have embedded the structural information (RPST) and the linguistic information (DSynTs) of the original process model into the intermediate representation RDT. Then, it is passed to the decoder phase.

\subsection{Text planning}\label{ssec:textPlanning}
The biggest gap between a model and a text is that a model contains sequential and concurrent semantics \cite{behavior}, while a text only contains sequential sentences. Thus, this step focuses on \textbf{transforming a non-linear model into its linear representations}.

\begin{figure}[hbtp]
  \centering
  \includegraphics[width=0.70\maxWidth]{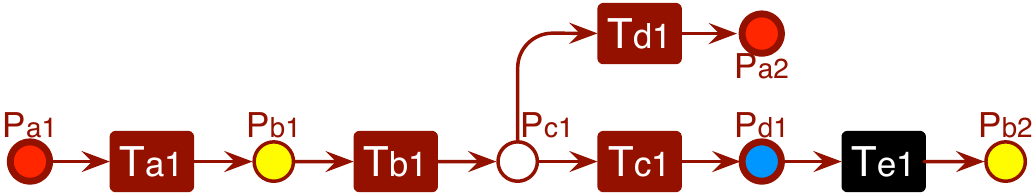}
  \caption{The CFP of $N_*$ ($\mathbb{N}_*$). The five colored places are shadow places. A shadow place is shown in same (different) color as its mutual (non-mutual) places.}
  \label{fig:rcfp}
  \end{figure}

In order to maintain behavior correctness, we first create the CFP of the original model because a CFP is a \textbf{complete and minimal behavior-unfolded graph} of the original model \cite{unfolding1, unfolding2, unfolding3}. Figure \ref{fig:rcfp} shows the CFP of $N_*$. According to Definition \ref{def:cfp}, $T_{d1}$ and $T_{e1}$ are two cut-off transitions, thus, no transitions follow them.

Besides, we introduce a basic concept: \textbf{\emph{shadow place}}. Shadow places ($\mathcal{SP}$) are those CFP places that are: 1) mutual with CFP boundary places or 2) mapped to the boundary places of the original model. 

\begin{example}
  In Figure \ref{fig:rcfp}, the five colored places are shadow places of $\mathbb{N}_*$ ($P_{a1},P_{b1},P_{a2},P_{d1},P_{b2}\in \mathcal{SP}(\mathbb{N}_*)$). Note that theyare mutual with the CFP boundary places, and $P_{d1}$ is mapped to the boundary places of the original model $N_*$ ($\hbar(P_{d1})=P_d$). Intuitively, a shadow place represents the repetition of a boundary place in the original model or its CFP.
\end{example}

\subsubsection{\textbf{Behavior Segment}}
Since we have obtained the behavior-unfolded graph, i.e., CFP, now, we define (behavior) segments which capture the \textbf{minimal behavioral characteristics of a CFP} to avoid state space explosion problem.

\begin{definition}[Behavior Segment] \label{def:segment}
  Given a net $N=(P,T,F)$ and its CFP $\mathbb{N}$, a behavior segment $\mathbb{S}=(P',T',F')$ is a connected sub-model of $\mathbb{N}$ satisfying:
  \begin{enumerate}[1)]
    \item $\bullet \mathbb{S} \bullet\subseteq\mathcal{SP}(\mathbb{N}) \wedge P' \backslash \bullet \mathbb{S} \bullet \cap \mathcal{SP}(\mathbb{N}) = \varnothing$, i.e., all boundary nodes are shadow places and all other places are not.
    \item If each place in $\hbar(\bullet\mathbb{S})$ contains one token, after firing all transitions in $\hbar(T')$, each place in $\hbar(\mathbb{S}\bullet)$ contains just one token while other places in $\hbar(\mathbb{N})$ are empty.
  \end{enumerate}
  \end{definition}

\begin{figure*}[hbtp]
  \centering
  \subfigure[The behavior segments of $\mathbb{N}_*$.]{
    \label{sfig:segments}
    \includegraphics[height=0.30\maxWidth]{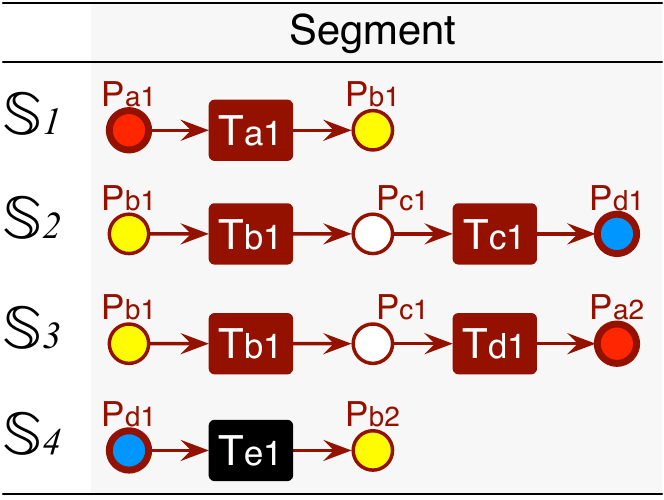}
  }
  \hspace{0.5cm}
  \subfigure[The partial behavior paths of $\mathbb{N}_*$.]{
    \label{sfig:paths}
    \includegraphics[height=0.30\maxWidth]{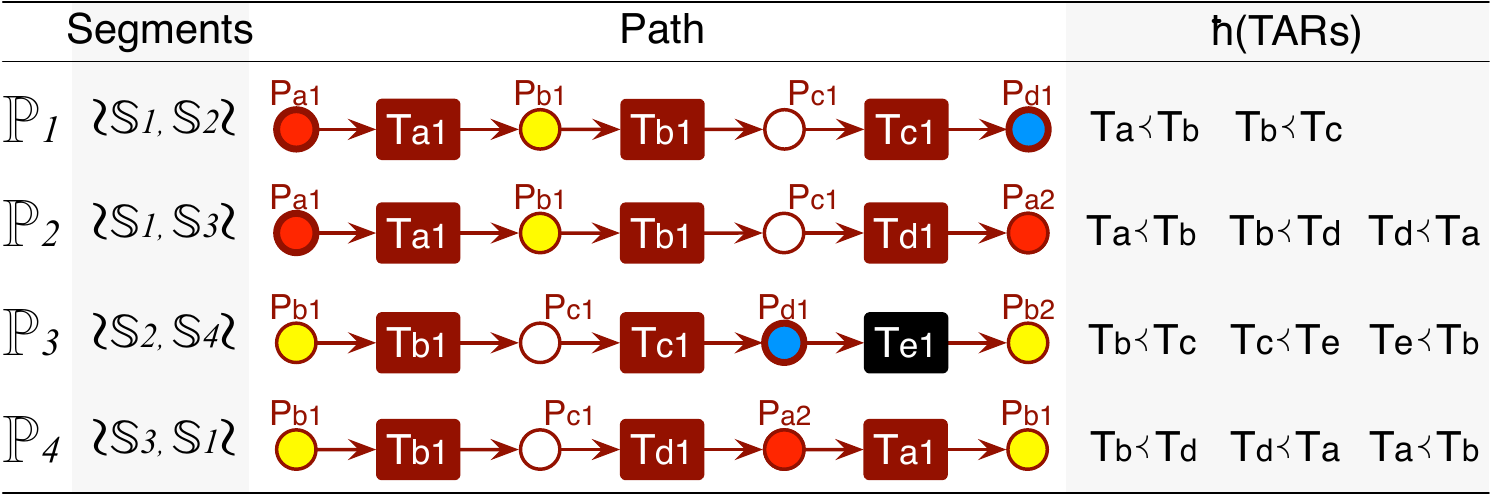}
  }
  \caption{The behavior segments and the partial behavior paths of $\mathbb{N}_*$.}
  \label{fig:segmentsPaths}
\end{figure*}

\begin{example}
  According to Definition \ref{def:segment}, if we put $\hbar(P_{a1})=P_a$ (in $N_*$) a token, $T_a$ (in $N_*$) can be fired, and after this firing, only $P_b$ (in $N_*$) contains a token. Therefore, the sub-model containing nodes $P_{a1}, T_{a1}, P_{b1}$ (in $\mathbb{N}_*$) and their adjacency arcs is a behavior segment. All behavior segments of $\mathbb{N}_*$ are shown in Figure \ref{sfig:segments} (careful readers might have realized that these four segments belong to sequential structures, i.e., all segments contain only SESE nodes. However, a behavior segment can be a non-sequential structure, i.e., containing multiple-incoming or multiple-outgoing nodes. For example, the behavior segment of a transition-bounded model is homogeneous to itself, containing four multiple-incoming or multiple-outgoing nodes).
\end{example}

\subsubsection{\textbf{Linking Rule}}
Behavior segments capture the minimal behavioral characteristics of a CFP. In order to portray the complete characteristics, we link these segments to obtain all possible behavior paths by applying the linking rule below.

\begin{definition}[Linking Rule] \label{def:linkingRule}
  For two segments $\mathbb{S}_i=(P_i,T_i,F_i)$ and $\mathbb{S}_j=(P_j,T_j,F_j)$, if $\hbar(\mathbb{S}_i\bullet)\supseteq \hbar(\bullet \mathbb{S}_j)$ we say they are linkable. If two places $p_i\in \mathbb{S}_i\bullet, p_j\in \bullet \mathbb{S}_j$ are mutual, we say $p_i$ is the joint place of $p_j$ denoted as $\mathcal{J}(p_j)=p_i$ where $\mathcal{J}$ is the joint function. If $n \notin \bullet \mathbb{S}_j$, $\mathcal{J}(n)=n$. The linked segment of two linkable segments $\wr \mathbb{S}_i, \mathbb{S}_j\wr=(P\wr,T\wr,F\wr)$ satisfies:

  \begin{enumerate}[1)]
    \item $P\wr=P_i\cup (P_j\backslash\bullet \mathbb{S}_j)$, i.e., the places of a linked segment consist of all places in $\mathbb{S}_i$ and all non-entry places in $\mathbb{S}_j$.
    \item $T\wr=T_i\cup T_j$, i.e., the transitions of a linked segment consist of all transitions in $\mathbb{S}_i$ and $\mathbb{S}_j$.
    \item $F\wr=\{\langle \mathcal{J}(u),\mathcal{J}(v) \rangle | \langle u,v \rangle \in F_i \cup F_j \}$, i.e., the arcs of a linked segment are the $\mathcal{J}$-replaced arcs of $\mathbb{S}_i$ and $\mathbb{S}_j$.
  \end{enumerate}
\end{definition}

\begin{figure}[hbtp]
  \centering
  \includegraphics[width=0.90\maxWidth]{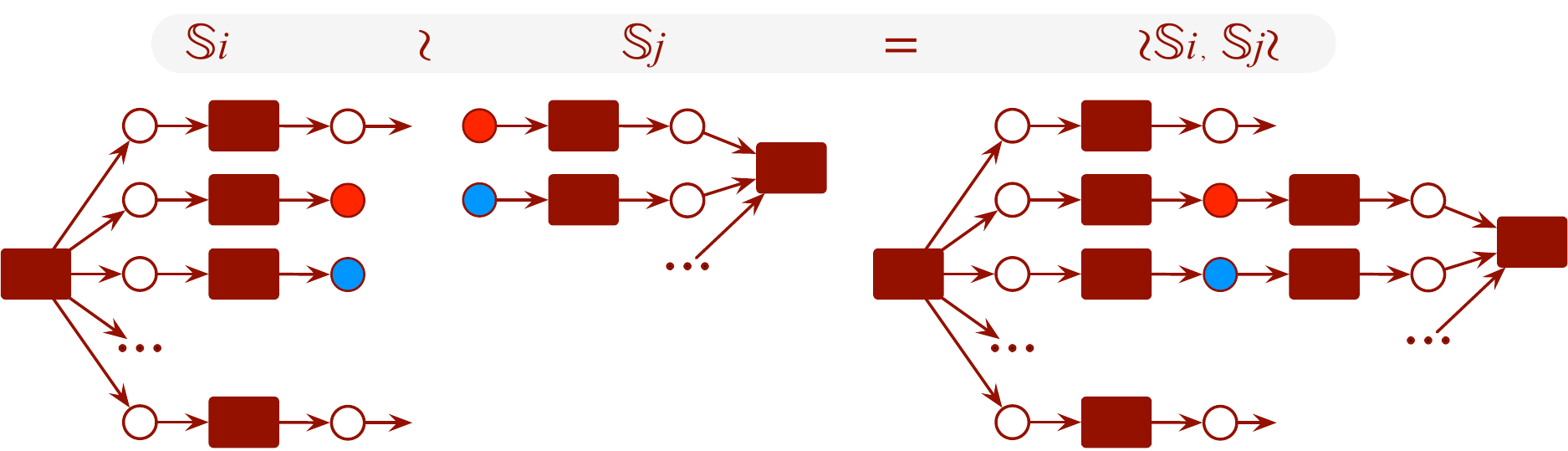}
  \caption{The graphical explanation of linking two segments $\mathbb{S}_i, \mathbb{S}_j$. The joint nodes are shown in red/blue color.}
  \label{fig:linkingRule}
  \end{figure}

Similarly, $\wr \mathbb{S}_1, \mathbb{S}_2, \cdots, \mathbb{S}_n\wr$ denotes the recursive linking of two segments  $\wr \mathbb{S}_1, \mathbb{S}_2, \cdots, \mathbb{S}_{n-1}\wr$ and $\mathbb{S}_n$. The graphical explanation of the linking rule is shown in Figure \ref{fig:linkingRule}.

\subsubsection{\textbf{Behavior Path}}
According to the linking rule, we can obtain all linked segments. However, a linked segment might involve infinite linking due to concurrent and loop behaviors \cite{behavior}. Hence, we apply truncation conditions to avoid infinite linking, which leads to the definition of a (behavior) path. Behavior paths capture \textbf{complete behavioral characteristics of a CFP}.

\begin{definition}[Behavior Path] \label{def:path}
  A segment $\mathbb{P}=\wr \mathbb{S}_1, \mathbb{S}_2, \cdots, \mathbb{S}_n\wr$ of $\mathbb{N}$ is a behavior path iff one of the following conditions holds:
  \begin{enumerate}[1)]
    \item $\bullet\mathbb{P}= \bullet \mathbb{N} \wedge \mathbb{P}\bullet \subseteq \mathbb{N}\bullet$, i.e.,  $\mathbb{P}$ starts from the entry of $\mathbb{N}$ and ends at one of the exits of $\mathbb{N}$.
    \item $\hbar(\bullet\mathbb{P})=\hbar(\mathbb{P}\bullet)$, i.e., $\mathbb{P}$ starts from a shadow node (set) and ends at this node (set), i.e., loop structure.
  \end{enumerate}
  \end{definition}

\begin{example}
Take Figure \ref{sfig:segments} as an example. Since $\hbar(\mathbb{S}_3\bullet)=\hbar(\{P_{a2}\})$\\=$\{P_{a}\}\supseteq \hbar(\bullet\mathbb{S}_1)=\hbar(\{P_{a1}\})=\{P_{a}\}$, it follows that $\mathbb{S}_3$ and $\mathbb{S}_1$ are linkable with $\mathcal{J}(p_{a1})=p_{a2}$, and $\hbar(\bullet \wr \mathbb{S}_3, \mathbb{S}_1\wr)=\hbar(\wr \mathbb{S}_3, \mathbb{S}_1\wr \bullet)=\{P_b\}$. Thus, the linked segment $\wr \mathbb{S}_3, \mathbb{S}_1\wr$ is a behavior path ($\mathbb{P}_4$ in Figure \ref{sfig:paths}). Partial behavior paths of $N_*$ are shown in Figure \ref{sfig:paths}.
\end{example}

\subsection{Sentence planning}
After extracting all behavior paths from a process model. Then, each path is recognized as a polygon component and then put into \emph{BePT} (a recursive algorithm). The endpoint is a non-decomposable trivial component, i.e., node. When encountering a gateway node (split or join node), the corresponding DSynT (a pre-defined language template) is retrieved from RDT or pre-defined XML-format files. When encountering a SESE node, the corresponding DSynT is extracted from the embedded RDT. After obtaining all DSynTs, the sentence planning phase is triggered.

Sentence planning sets out to generate a sentence for each node. The main idea here is to utilize a DSynT to create a NL sentence \cite{realpro, dsyn, Hen}. The generation task is divided into two levels: template sentence and activity sentence generation.

\begin{enumerate}[$\bullet$]
  \item \textbf{Template sentences} focus on describing the behavioral information related to the non-terminal RPST nodes. We provide 32 language template DSynTs (including split, join, dead transition, deadlock \cite{processMining} etc,) to represent corresponding semantic(s). The choice of a template depends on three parameters \cite{Leo, Hen, Goun}: 1) the existence of a gateway label; 2) the gateway type; 3) the number of outgoing arcs. For instance, for a place with multiple outgoing arcs, the corresponding template sentence ``\emph{One of the branches is executed}'' will be retrieved.

  \item \textbf{Activity sentences} focus on describing a single activity related to the terminal (leaf) RPST nodes. RDT representation has embedded all DSynT messages; thus, for each activity, we can directly access its DSynT from RDT.
  \end{enumerate}

After preparing all DSynTs in the text planning phase, we employ three steps to optimize the expression before the final generation:
\begin{enumerate}[1)]
  \item Checking whether each DSynT lacks necessary grammar meta-information to guarantee its grammatical correctness. 
  \item Pruning redundant TARs to ensure that the selected TARs will not be repeated (Pruning Rule). For example, $T_a \prec T_b$ derived by $\mathbb{P}_2$ or $\mathbb{P}_4$ in Figure \ref{sfig:paths} is a redundant TAR because it has been concluded in $\mathbb{P}_1$.
  \item Refining the DSynT messages containing the same linguistic component between two consecutive sentences and making use of three aggregation strategies: role aggregation, action aggregation and object aggregation \cite{Hen, Goun}.
  \end{enumerate}

After expression optimization, we employ the DSynT-based realizer RealPro \cite{realpro} to realize sentence generation. RealPro requires a DSynT as input and outputs a grammatically correct sentence \cite{dsyn}. In a loop, every DSynT is passed to the realizer. The resulting NL sentence is then added to the final output text. After all sentences have been generated, the final text is presented to the end user.

\begin{example} \label{exm:text}
The generated text of $N_*$ in Figure \ref{fig:rigid} is as follows (other state-of-the-art methods cannot handle this model):
\begin{mdframed}[style=myboxstyle,frametitle={}]
  \begin{enumerate}[1)]
    \item $\bullet$ The following main branches are executed:
    \item \hspace{1em} $\bullet$ The experimenter extracts the genes. Then, he sequences the DNA. Subsequently, the experimenter records the data.
    \item $\bullet$ Attention, there are two loops which may conditionally occur:
    \item \hspace{1em} $\bullet$ After sequencing DNA, the experimenter can also remove impurities if it is not clean. Then, he continues extracting genes.
    \item \hspace{1em} $\bullet$ After recording the data, there is a series of activities that need to be finished before DNA sequencing:
    \item \hspace{3em} $\bullet$ ***
    % \item $\bullet$ Once the data is enough, the process ends.
  \end{enumerate}
\end{mdframed}

Template sentences (1, 3, 7) describe where the process starts, splits, joins and ends. Activity sentences (2, 4, 5) describe each sorted behavior path. The paragraph placeholder (6) can be flexibly replaced according to the sub-text of the simplified component $T_e$. We can see that \emph{BePT} first describes the main path (``\emph{\underline{$T_a$}$\rightarrow$\underline{$T_b$}$\rightarrow$\underline{$T_c$}}'') before two possible loops (``\emph{\underline{$T_a$}$\rightarrow$\underline{$T_b$}$\rightarrow$\underline{$T_d$}}'', ``\emph{\underline{$T_c$}$\rightarrow$\underline{$T_e$}$\rightarrow$\underline{$T_b$}}''). These three paragraphs of the generated text correspond to three correct firing sequences of the original model, the generated text contains just enough descriptions to reproduce the original model without redundant descriptions.
\end{example}

\begin{table*}[hbtp]
  \setlength{\abovecaptionskip}{0pt}
  \setlength{\belowcaptionskip}{0pt}
  \small
  % \scriptsize
  \centering
  \caption{Statistics of the evaluation datasets. \emph{N}=Total number of models per source; \emph{SMR}=The ratio of structured models to all models; \emph{min}=The minimum value per source; \emph{ave}=The average value per source; \emph{max}=The maximum value per source.}
  \label{tab:datasets}
  \rowcolors{3}{grayColor}{whiteColor}
  \begin{tabular}{c|c|r|r|rrr|rrr|rrr|rrr}
    \hline
    \multirow{2}*{\textbf{Source}} & \multirow{2}*{\textbf{Type}} & \multirow{2}*{\textbf{N}} & \multirow{2}*{\textbf{\emph{SMR} $\downarrow$}} & \multicolumn{3}{c|}{\textbf{Place}} & \multicolumn{3}{c|}{\textbf{Transition}} & \multicolumn{3}{c|}{\textbf{Arc}} & \multicolumn{3}{c}{\textbf{RPST depth}}\\
    % \cmidrule(lr){5-7}\cmidrule(lr){8-10}\cmidrule(lr){11-13}\cmidrule(lr){14-16}
    &  &  &  & \emph{min} & \emph{ave} & \emph{max} &  \emph{min} & \emph{ave} & \emph{max} & \emph{min} & \emph{ave} & \emph{max} & \emph{min} & \emph{ave} & \emph{max} \\
    \hline
    SAP & Industry & 72  & \cellcolor{redColor!100}100.00\% & 2 & 3.95 & 13 & 1 & 3.12 & 12 & 2 & 6.75 & 24 & 1 & 1.85 & 5 \\
    DG  & Industry & 38  & \cellcolor{redColor!85}94.74\%   & 3 & 7.65 & 22 & 2 & 7.85 & 17 & 4 & 16.02 & 44 & 1 & 2.55 & 7 \\
    TC  & Industry & 49  & \cellcolor{redColor!70}81.63\%   & 6 & 10.10 & 17 & 6 & 10.62 & 19 & 14 & 21.87 & 38 & 1 & 3.92 & 7 \\
    SPM & Academic & 14  & \cellcolor{redColor!55}57.00\%   & 2 & 7.28 & 12 & 1 & 7.40 & 15 & 2 & 15.49 & 30 & 1 & 2.93 & 5 \\
    IBM & Industry & 142 & \cellcolor{redColor!40}53.00\%   & 4 & 39.00 & 217 & 3 & 26.46 & 145 & 6 & 79.84 & 456 & 1 & 5.21 & 12 \\
    GPM & Academic & 36  & \cellcolor{redColor!25}42.00\%   & 4 & 11.15 & 19 & 3 & 11.55 & 24 & 6 & 24.92 & 48 & 1 & 3.22 & 5 \\
    BAI & Academic & 38  & \cellcolor{redColor!10}28.95\%   & 4 & 10.54 & 21 & 2 & 9.93 & 24 & 6 & 22.92 & 49 & 1 & 3.24 & 5 \\
    \hline
    \end{tabular}
  \end{table*}

\subsection{Property Analysis} \label{ssec:property}
We emphasize BePT's three strong properties - correctness, completeness and minimality. Specifically, given a net system $S=(N,M)$ and its TAR set $\mathcal{T}(S)$. The behavior path set $\mathcal{P}$ of $\mathbb{N}$ by the linking rule (Definition \ref{def:linkingRule}) satisfies: 1) \textbf{behavior correctness}, $\forall \mathbb{P} \in \mathcal{P} \Rightarrow \mathcal{T}(\mathbb{P}) \subseteq \mathcal{T}(S)$; 2) \textbf{behavior completeness}, $\forall \tau \in \mathcal{T}(S) \Rightarrow \exists \mathbb{P}\in\mathcal{P}, \tau \in \mathcal{T}(\mathbb{P})$; 3) \textbf{description minimality}, each TAR $\mathcal{T}(S)$ by the pruning rule is described only once in the final text. Please see Appendices A, B and C for detailed proofs.

\section{Evaluation} \label{sec:evaluation}

We have conducted extensive qualitative and quantitative experiments. In this section, we report the experimental results to answer the following research questions:

\begin{enumerate}[\textbf{RQ}1]
  \item \textbf{Capability}: Can \emph{BePT} handle more complex model patterns than existing techniques?
  \item \textbf{Detailedness}: How much information does \emph{BePT} express?
  \item \textbf{Consistency}: Is \emph{BePT} text consistent to the original model?
  \item \textbf{Understandability}: Is \emph{BePT} text easy to understand?
  \item \textbf{Reproducibility}: Can the original model be reproduced only from its generated text?
\end{enumerate}

\subsection{Experimental Setup} \label{ssec:setup}
In this part, we describe our experimental datasets, the baselines and the experiment settings.

\subsubsection{\textbf{Datasets}} \label{sssec:datasets}
We collected and tested on seven publicly accessible datasets: SAP, DG, TC, SPM, IBM, GPM, BAI \cite{Leo,Hen,Goun,cims}. Among them, SAP, DG, TC, IBM are from industry (enterprises etc,) and SPM, GPM, BAI are from academic areas (literature, online tutorials, books etc,). The characteristics of the seven datasets are summarized in Table \ref{tab:datasets} (sorted by the decreasing ratio of structured models $\bm{SMR}$). There are a total of 389 process models consisting of real-life enterprise models (87.15\%) and synthetic models (12.85\%). The number of transitions varies from 1 to 145 and the depth of RPSTs varies from 1 to 12. The statistical data is fully skewed due to the different areas, amounts and model structures.

\subsubsection{\textbf{Baseline Methods}} \label{sssec:baselines}
We compared our proposed process translator \emph{BePT} with the following three state-of-the-art methods:
\begin{enumerate}[$\bullet$]
  \item \emph{\textbf{Leo}} \cite{Leo}. It is the first structure-based method focusing mainly on structured components: trivial, bond and polygon.
  \item \emph{\textbf{Hen}} \cite{Hen}. It is the extended version of Leo focusing mainly on rigid components with longest-first strategy.
  \item \emph{\textbf{Goun}} \cite{Goun}. It is a state-of-the-art structured-based method focusing mainly on unfolding model structure without considering its behaviors.
\end{enumerate}

\subsubsection{\textbf{Parameter Settings}} \label{sssec:settings}
We implemented \emph{BePT} based on jBPT\footnote{\url{https://code.google.com/archive/p/jbpt/}}. An easy-to-use version of \emph{BePT} is also publicly available\footnote{\url{https://github.com/qianc62/BePT}}. We include an editable parameter for defining the size of a paragraph and predefine this parameter with a value of 75 words. Once this threshold is reached, we use a change of the performing role or an intermediate activity as an indicator and respectively introduce a new paragraph. Besides, we use the default language grammar style of subject-predicate-object and object-be-predicated-by-subject to express a sentence \cite{Leo, Hen, Goun}. Finally, we set all parameters to valid for all methods, i.e., to generate intact textual descriptions without any reduction.

\subsection{Results} \label{ssec:results}

\subsubsection{\textbf{Capability (RQ1)}} \ \\
As discussed earlier, a rigid is a region that captures an arbitrary model structure. Thus, these seven datasets are representative enough as the $SMR$ varies from 100\% (structured models) to 28.95\% (unstructured complex models). We analyzed and compared all process models. Table \ref{tab:capability} reports their handling capabilities w.r.t some representative complex patterns \cite{Goun, unstructuredModel}.

\begin{table}[htbp]
  \setlength{\abovecaptionskip}{0pt}
  \setlength{\belowcaptionskip}{0pt}
  \small
  \centering
  \caption{The handling capabilities of four P2T methods w.r.t. some representative patterns.}
  \label{tab:capability}
  \begin{tabular}{c|c|c|c|c|c}
    \hline
    \textbf{Type} & \textbf{Pattern} & \textbf{\emph{Leo}} & \textbf{\emph{Hen}} & \textbf{\emph{Goun}} & \textbf{\emph{BePT}} \\
    \hline
    \multirow{5}*{T, B, P} & \cellcolor{grayColor}Trivial & \cellcolor{grayColor}\color{redColor}\checkmark & \cellcolor{grayColor}\color{redColor}\checkmark & \cellcolor{grayColor}\color{redColor}\checkmark & \cellcolor{grayColor}\color{redColor}\checkmark \\
    & Polygon & \color{redColor}\checkmark & \color{redColor}\checkmark & \color{redColor}\checkmark & \color{redColor}\checkmark \\
    & \cellcolor{grayColor}Easy Bond & \cellcolor{grayColor}\color{redColor}\checkmark & \cellcolor{grayColor}\color{redColor}\checkmark & \cellcolor{grayColor}\color{redColor}\checkmark & \cellcolor{grayColor}\color{redColor}\checkmark \\
    & Easy Loop & \color{redColor}\checkmark & \color{redColor}\checkmark & \color{redColor}\checkmark & \color{redColor}\checkmark \\
    & \cellcolor{grayColor}Unsymmetrical Bond & \cellcolor{grayColor} & \cellcolor{grayColor} & \cellcolor{grayColor} & \cellcolor{grayColor}\color{redColor}\checkmark \\
    \hline
    \multirow{7}*{R} & Place Rigid & & \color{redColor}\checkmark & \color{redColor}\checkmark & \color{redColor}\checkmark \\
    & \cellcolor{grayColor}Transition Rigid & \cellcolor{grayColor} & \cellcolor{grayColor} & \cellcolor{grayColor} & \cellcolor{grayColor}\color{redColor}\checkmark \\
    & Mix Rigid & & & & \color{redColor}\checkmark \\
    & \cellcolor{grayColor}Intersectant Loop & \cellcolor{grayColor} & \cellcolor{grayColor} & \cellcolor{grayColor} & \cellcolor{grayColor}\color{redColor}\checkmark \\
    & Non-free-choice Construct & & & \color{redColor}\checkmark & \color{redColor}\checkmark \\
    & \cellcolor{grayColor}Invisible or Duplicated Task & \cellcolor{grayColor} & \cellcolor{grayColor} & \cellcolor{grayColor}\color{redColor}\checkmark & \cellcolor{grayColor}\color{redColor}\checkmark \\
    & Multi-layered Embedded & & & \color{redColor}\checkmark & \color{redColor}\checkmark \\
    \hline
    \multirow{2}*{Extra} & \cellcolor{grayColor}Modeling Information & \cellcolor{grayColor} & \cellcolor{grayColor} & \cellcolor{grayColor} & \cellcolor{grayColor}\color{redColor}\checkmark \\
    & Multi-layered Paragraph & & & \color{redColor}\checkmark & \color{redColor}\checkmark \\
    \hline
    \multicolumn{2}{c|}{Total} & \cellcolor{redColor!25}4 & \cellcolor{redColor!40}5 & \cellcolor{redColor!70}9 & \cellcolor{redColor!100}14 \\
    \hline
    \end{tabular}
\end{table}

First, we can see that \emph{BePT} shows the best handling capabilities. Among the 14 patterns, \emph{BePT} can handle them all, which is better than Goun that can handle 9 patterns. Second, all four methods can handle structured models well, while \emph{Goun} and \emph{BePT} can handle unstructured models, and \emph{BePT} can even further provide extra helpful messages. Third, the R and the Extra parts show that \emph{BePT} can handle rigids of arbitrary complexity even if the model is unsymmetrical, non-free-choice or multi-layered. From these results, we can conclude that the behavior-based method \emph{BePT} is sufficiently powerful to address complex structures.

\subsubsection{\textbf{Detailedness (RQ2)}} \ \\
In the sentence planning phase, \emph{BePT} checks the grammatical correctness of each DSynT so that the generated text can accord with correct English grammar. Here, instead of comparing the grammatical correctness, we summarize the structural characteristics of all generated texts in Table \ref{tab:detailedness}.

\begin{table}[t]
\small
\centering
\caption{Average number of words and sentences per text. Red numbers denote the maximum and green numbers denote the minimum per dataset.}
\label{tab:detailedness}
\rowcolors{3}{grayColor}{whiteColor}
\begin{tabular}{c|rrrr|rrrr}
\hline
\multirow{2}*{} & \multicolumn{4}{c|}{\textbf{Words/Text}} & \multicolumn{4}{c}{\textbf{Sentences/Text}} \\
% \cmidrule(lr){2-3}\cmidrule(lr){4-5}\cmidrule(lr){6-7}\cmidrule(lr){8-9}
& Leo & Hen & Goun & \emph{BePT} & Leo & Hen & Goun & \emph{BePT} \\
\hline
SAP   & \color{greenColor} 38.0  & \color{greenColor} 38.0  & \color{redColor} 38.1  & \color{redColor} 38.1  & \color{greenColor} 6.0  & \color{greenColor} 6.0  & \color{redColor} 6.2  & \color{redColor} 6.2  \\
DG    & \color{greenColor} 74.0  & 79.7  & 79.6  & \color{redColor} 85.3  & \color{greenColor} 13.0 & 15.0 & 15.0 & \color{redColor} 15.7 \\
TC    & \color{greenColor} 99.2  & 110.8 & 112.4 & \color{redColor} 135.0 & \color{greenColor} 12.2 & 15.5 & 15.7 & \color{redColor} 18.7 \\
SPM   & \color{greenColor} 41.5  & 54.1  & 55.6  & \color{redColor} 100.9 & \color{greenColor} 5.8  & 7.9  & 8.1  & \color{redColor} 14.1 \\
IBM   & \color{greenColor} 140.2 & 180.7 & 182.9 & \color{redColor} 191.9 & \color{greenColor} 74.2 & 80.7 & 81.7 & \color{redColor} 86.2 \\
GPM   & \color{greenColor} 38.3  & 50.8  & 53.8  & \color{redColor} 147.0 & \color{greenColor} 6.2  & 7.5  & 7.9  & \color{redColor} 16.2 \\
BAI   & \color{greenColor} 25.7  & 31.7  & 32.6  & \color{redColor} 111.3 & \color{greenColor} 2.7  & 4.4  & 4.6  & \color{redColor} 15.7 \\
\hline
Total & \color{greenColor} 66.7  & 78.0  & 79.3  & \color{redColor} 115.3 & \color{greenColor} 17.2 & 19.6 & 19.9 & \color{redColor} 22.3 \\
\hline
\end{tabular}
\end{table}

A general observation is that \emph{BePT} texts are longer than the other texts. Leo, Hen, and Goun texts contain an average of 66.7, 78.0 and 79.3 word length and 17.2, 19.6, 19.9 sentence length respectively, while \emph{BePT} texts include an average of 115.3 words and 22.3 sentences. However, this does not imply that \emph{BePT} texts are verbose, using longer sentences to describe the same content. Rather, Leo, Hen, Goun ignore some modeling-level messages related to soundness and safety \cite{verification, processMining}, but \emph{BePT} supplements them. Therefore, we conclude that \emph{BePT} generates more detailed messages to provide additional useful information. Of course, while all parameters are set to be valid in this experiment, \emph{BePT} is actually configurable, i.e., users can set parameters to determine whether to generate these complementary details or not.

\subsubsection{\textbf{Consistency (RQ3)}} \ \\
A generally held belief is that the hierarchical organization of texts will hugely influence readability since paragraph indentation can reflect the number of components, the modeling depth of each activity, etc. Considering the generated text of the running example (Example \ref{exm:text}), if the text contains no paragraph indentation, i.e., each paragraph starts from the bullet point "$\bullet$", it will be much harder to fully reproduce the model semantics \cite{Hen, Goun}.

In this part, we consider the detection of structural consistency between a process model and its corresponding textual descriptions. This task requires an alignment of a model and a text, i.e., activities in the texts need to be related to model elements and vice versa \cite{alignment, consistency}. For an activity $T$, its modeling depth $md(T)$ is the RPST depth of $T$, and its description depth $dd(T)$ is how deep it is indented in the text. For the activity set of a model, the modeling depth distribution is denoted as $\mathcal{X}=[md(T_1),md(T_2),...,md(T_n)]$ and the description depth distribution is denoted as $\mathcal{Y}=[dd(T_1),dd(T_2),...,dd(T_n)]$. We employ a correlation coefficient to evaluate the consistency between the two distributions of $\mathcal{X}$ and $\mathcal{Y}$ as follows:
\begin{equation}
  \rho(\mathcal{X},\mathcal{Y})=\frac{E[(\mathcal{X}-E\mathcal{X})(\mathcal{Y}-E\mathcal{Y})]}{\sqrt{D\mathcal{X} \cdot D\mathcal{Y}}}\in [-1.0,1.0]
\end{equation}
where $E$ is the expectation function and $D$ is the variance function. The value of the $\rho(\mathcal{X},\mathcal{Y})$ function ranges from -1.0 (negatively related) to 1.0 (positively related).

\begin{figure}[htbp]
  \centering
  \includegraphics[width=0.70\maxWidth]{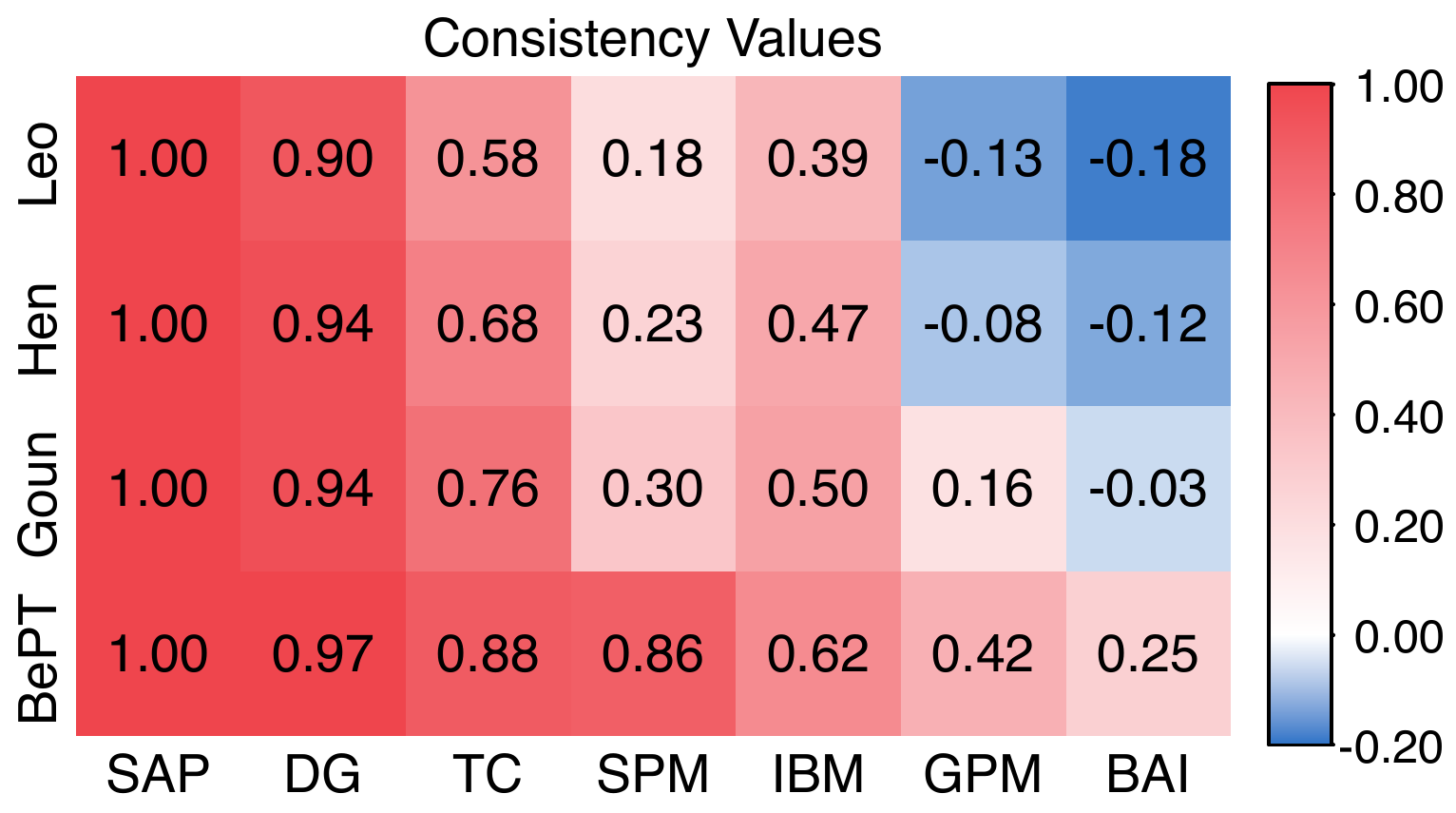}
  \caption{The consistency distribution. The red color denotes the positive coefficient while the blue color denotes the negative coefficient.}
  \label{fig:consistency}
\end{figure}

\begin{figure*}[hbtp]
  \centering
  \subfigure[Information gain line.]{
    \label{sfig:line}
    \includegraphics[height=0.30\maxWidth]{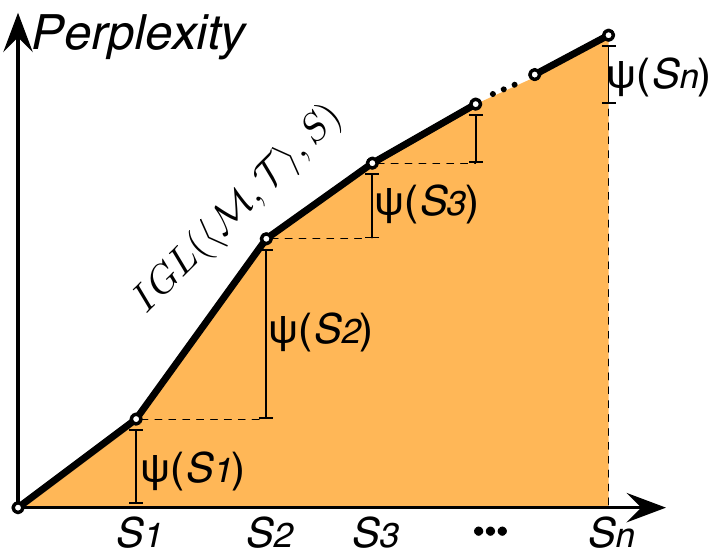}
  }
  \hspace{0.4cm}
  \subfigure[Perplexity distributions on all datasets.]{
    \label{sfig:perplextity}
    \includegraphics[height=0.35\maxWidth]{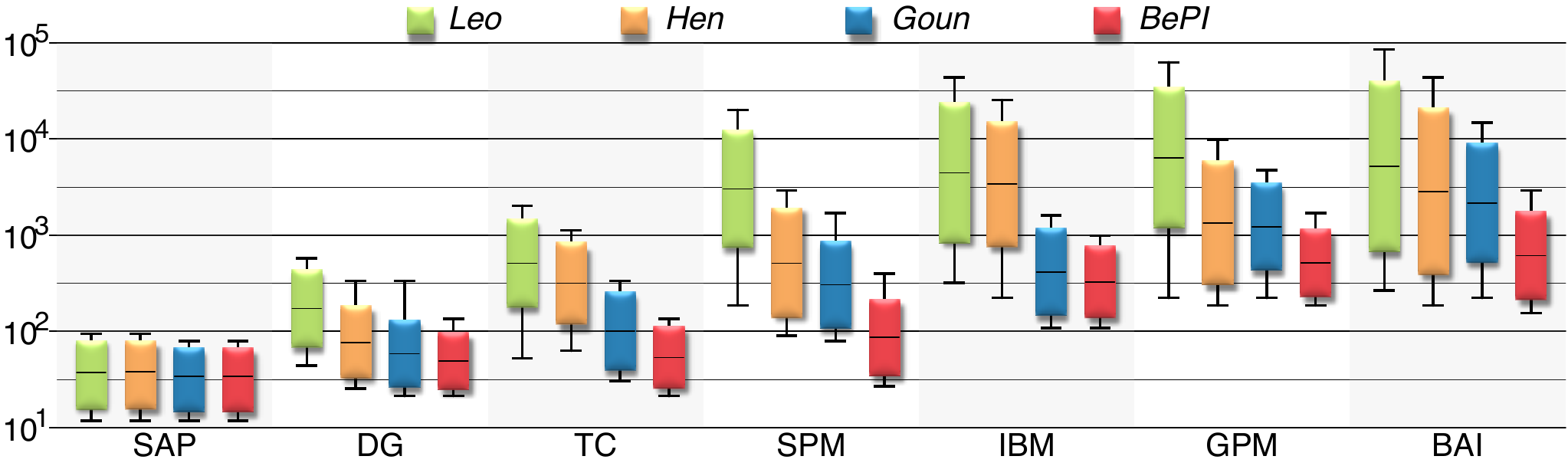}
  }
  \caption{The graphical representation of information gain line and the perplexity distributions.}
  \end{figure*}
  
Figure \ref{fig:consistency} shows the consistency results of the four P2T methods. First, \emph{BePT} obtains the highest consistency value in every dataset, meaning that \emph{BePT} positively follows the depth distribution of original models to the maximum extent. Notice that all methods obtain 1.00 consistency on the SAP dataset since all SAP models are structured. However, on the SPM dataset, \emph{BePT} achieves 0.86 consistency, while the other methods are only at around 0.25. The main reason is that SPM contains plenty of close-to-structured rigids, which directly reflects the other methods' drawbacks. Second, with lower $SMR$, the consistency performance rapidly decreases. The most obvious updates occur in GPM and BAI where Leo, Hen and Goun even produce negative coefficient values, which demonstrates that they negatively relate the distribution of the original models even causing the opposite distribution, while \emph{BePT} obtains 0.42 and 0.25 which shows that \emph{BePT} is still positively related even while facing unstructured situations. Hence, we conclude that \emph{BePT} texts conform better to the original models.

\subsubsection{\textbf{Understandability (RQ4)}} \ \\
In this section, we discuss the perplexity that reflects the textual understandability. It quantifies ``how hard to understand'' a model-text pair. This information entropy-based metric \cite{entropy1} is inspired from the natural language processing techniques \cite{perplexity}.

Consider a model-text pair $\langle \mathcal{M}, \mathcal{T} \rangle$ in which the text $\mathcal{T}$ consists of a sequence of paragraphs $\langle S_1,S_2,\cdots,S_n\rangle$. $\psi(S_i)$ denotes the information gain of paragraph $S_i$:
\begin{equation}
  \psi(S_i)=e^{|T'|log_2|T'|} \cdot |T'| \cdot |U'|
\end{equation}
where $T'$ is the described activity set and $U'$ is the neglected activity set. This formula employs information entropy $|T'|log_2|T'|$ to describe the confusion of all activities in a paragraph. Its exponent value has the same magnitude of $|T'|$. We notice that if any activity cannot be generated in the text, the text system should reduce the understandability value with the original model, i.e., improve the perplexity of the text system; hence, it multiplies by $|U'|$.

When describing a single paragraph $S_1$, the information gain \cite{dataMining} of the text system is $\psi(S_1)$. After describing paragraph $S_2$, the information gain changes to $\psi(S_1)+\psi(S_2)$. Similarly, after describing all paragraphs, the information gain is $\Sigma_{i=1}^n\psi(S_i)$. These values are mapped to $n$ points $(i,\Sigma_{k=1}^{i}\psi(S_k))_{i=1}^n$ shown in Figure \ref{sfig:line}. We call the broken line linking all points the \emph{information gain line} $IGL(\langle \mathcal{M}, \mathcal{T} \rangle,S)$. Then, we can define the perplexity of the text system $\mathcal{T}$ (the integral over all sentence perplexities): 

\begin{equation}
  perplexty(\langle \mathcal{M}, \mathcal{T} \rangle) = \int_{0}^{n}IGL(\langle \mathcal{M}, \mathcal{T} \rangle, s)ds, s\in \mathbb{R}
\end{equation}

$IGL(S)$ intuitively measures whether the model-text pair system is understandable where a lower perplexity implies a higher understandability. We calculated this metric for each dataset and reported the results.

Figure \ref{sfig:perplextity} shows the perplexity results. We can see that \emph{BePT} achieves the lowest perplexity in all datasets, i.e., best understandability. On average, the perplexity has been reduced from $10^{2.74}$ to $10^{0.98}$. This results also show that the perplexity trend is Leo $\ge$ Hen $\ge$ Goun $\ge$ BePT, i.e., the understandability trend is Leo $\le$ Hen $\le$ Goun $\le$ BePT.

\subsubsection{\textbf{Reproducibility (RQ5)}} \ \\
This part evaluates the reproducibility of the generated text, i.e., could the original model be reproduced from the generated text?

For each model-text pair ($\mathcal{M}, \mathcal{T}$), we manually back-translate (extract) the process model from the generated text and compare the elements between the original and the extracted models. All back-translators are provided only the generated texts without them knowing any information of the original models. They reproduce the original models from the texts according to their own understanding. After translation, we evaluate the structural and behavioral reproducibility between the original model and the extracted one. If an isomorphic model $\mathcal{M}$ can be reproduced, we can believe that the text $\mathcal{T}$ contains enough information to reproduce the original model, i.e., excellent reproducibility. We evaluate the P2T performance using the $F_1$ measure (the harmonic average of recall and precision) which is inspired by the data mining field \cite{dataMining}:
\begin{equation}
  F_1 = \frac{(1+\beta^2)\cdot precision \cdot recall}{\beta^2 \cdot precision + recall} \in [0.0,1.0]
\end{equation}
where $\beta$ is the balance weight. In our experiments, equal weights ($\beta=1.0$) are assigned to balance recall and precision. The higher the $F_1$ is, the better the reproducibility is. %$\theta_x(H)$ is an extraction function to extract the $x$ set from $H$ (e.g., $\theta_{place}(\mathcal{T})$ denotes the place set in the generated text $\mathcal{T}$).

\begin{figure}[t]
  \centering
  \subfigure[$F_1$ score on places]{
    \label{sfig:placeF}
    \includegraphics[width=0.40\maxWidth]{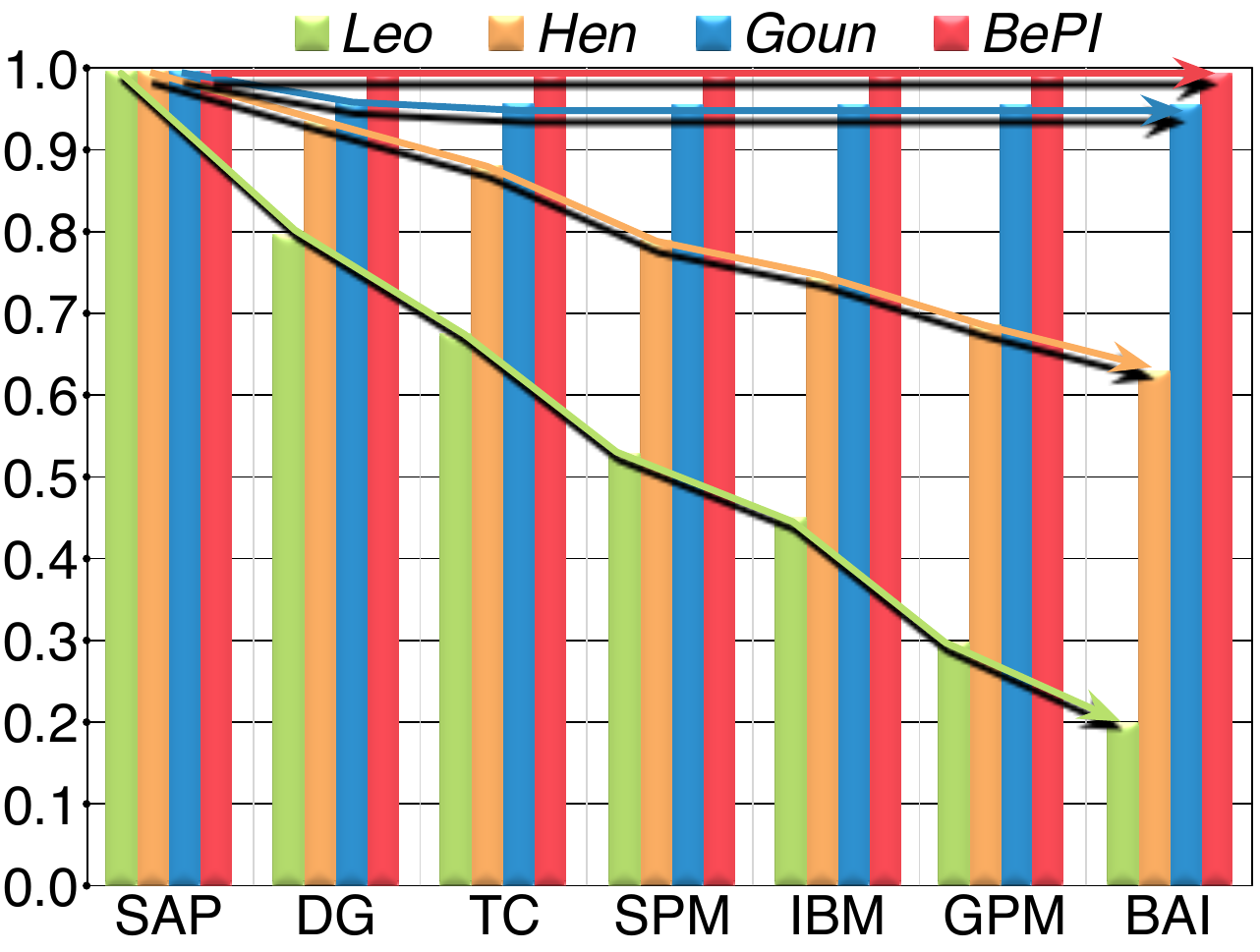}
  }
  \subfigure[$F_1$ score on transitions]{
    \label{sfig:transitionF}
    \includegraphics[width=0.40\maxWidth]{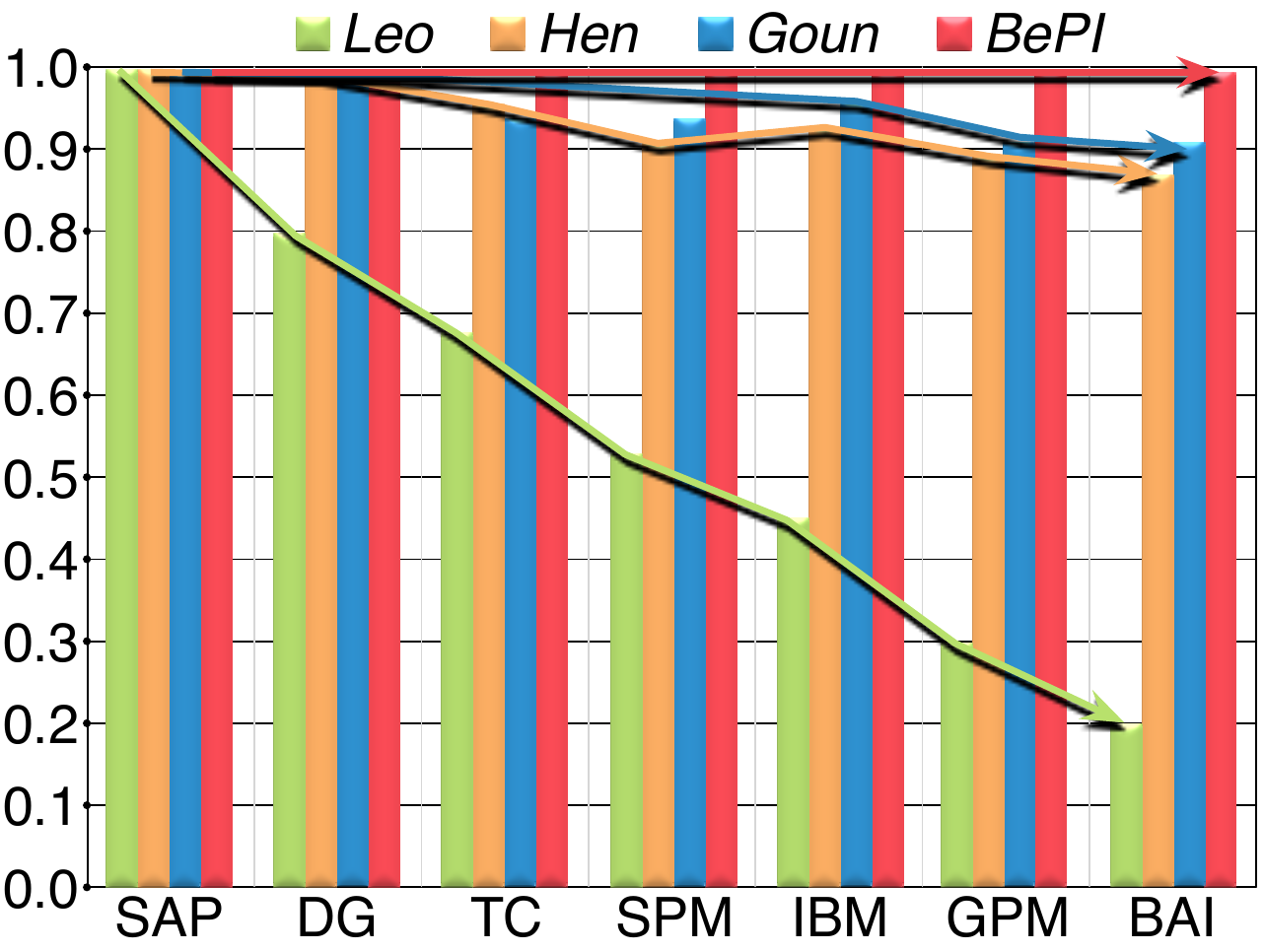}
  }
  \subfigure[$F_1$ score on gateways]{
    \label{sfig:gatewayF}
    \includegraphics[width=0.40\maxWidth]{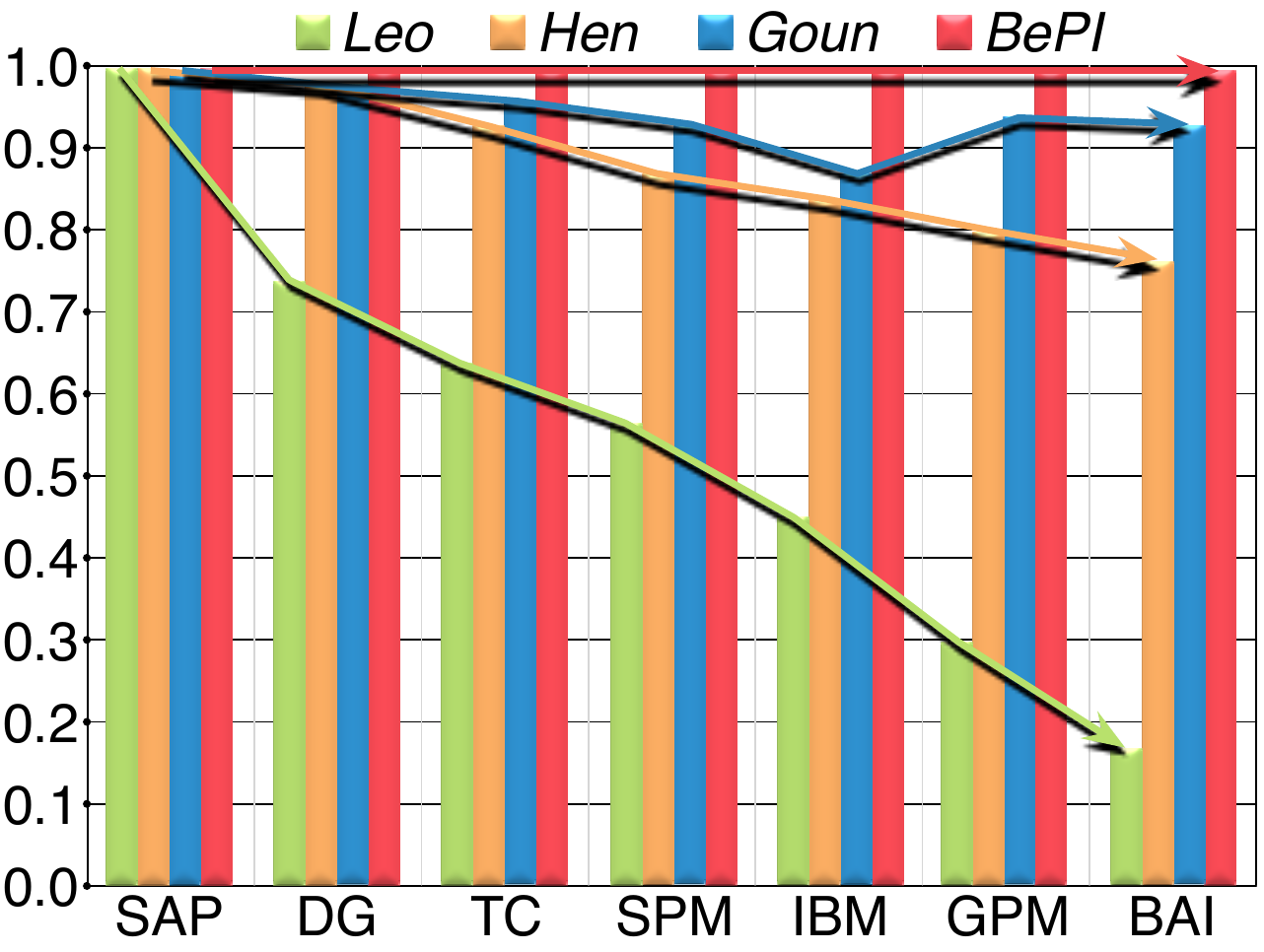}
  }
  \subfigure[$F_1$ score on all elements]{
    \label{sfig:elementF}
    \includegraphics[width=0.40\maxWidth]{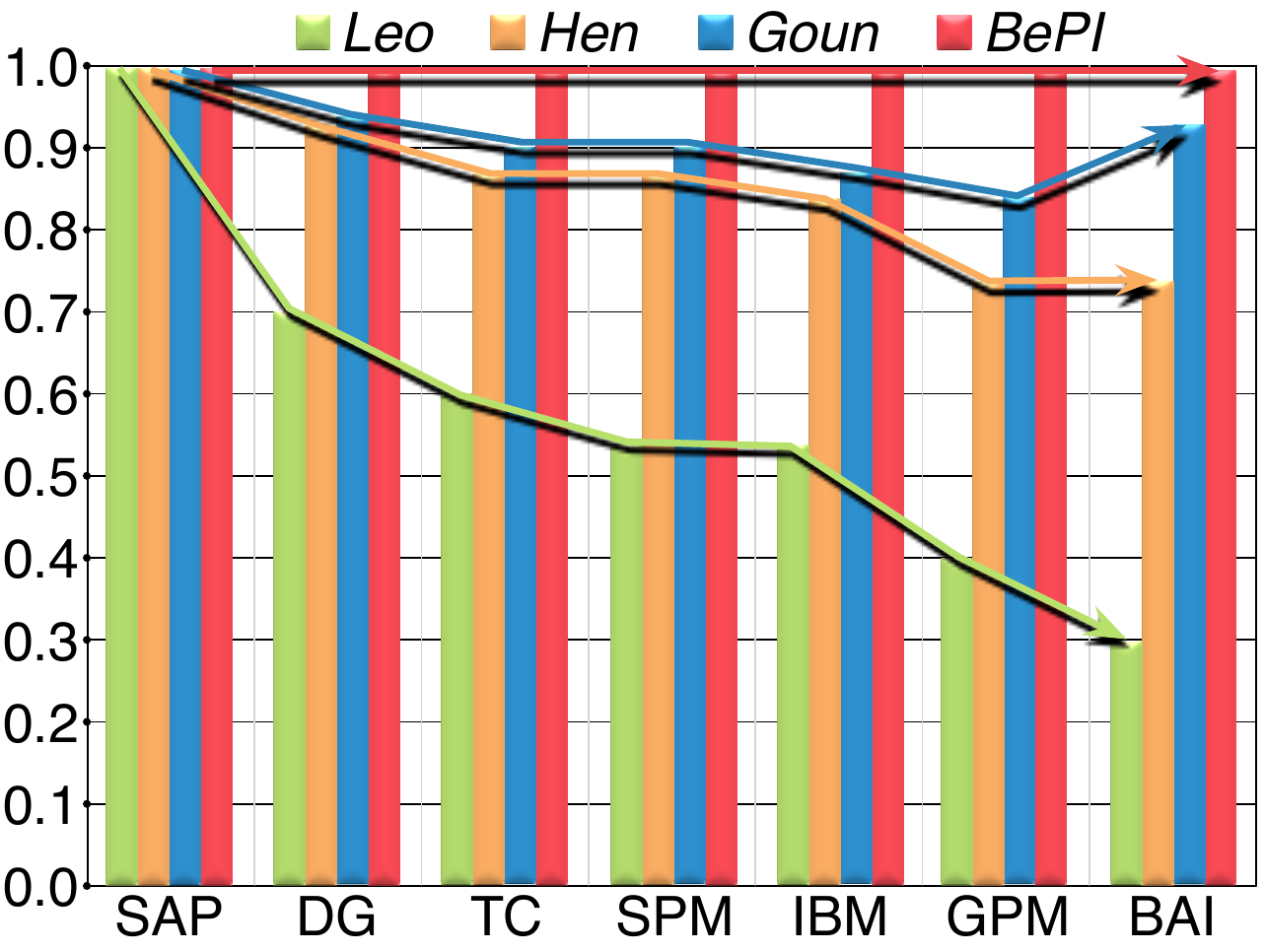}
  }
  \caption{The $F_1$ measures on structural dimensions.}
  \label{fig:structureMeasures}
  \end{figure}

\textbf{Structural Reproducibility.} Figure \ref{fig:structureMeasures} shows the results of four dimensions (\emph{place, transition, gateway, element}). First, we can see that the $F_1$ value of the four methods falls from 100\% to a lower value w.r.t. decreasing $SMR$. For GPM and BAI datasets, Leo achieves only around 40\%. The low-value cases significantly affect the ability to understand or reproduce the original model, and it reflects the general risk that humans may miss elements when describing a model, i.e., they lose around 60\% information. Still, Hen achieves around 90\% while \emph{BePT} hits 100\%, i.e., Goun and \emph{BePT} lose least information. We can conclude that, among the four P2T methods, \emph{BePT} achieves the highest reproducibility, followed by Goun and then Hen. The structural reproducibility performance also shows the trend, Leo $\le$ Hen $\le$ Goun $\le$ BePT.

\textbf{Behavioral Reproducibility.}
Behavioral reproducibility aims to evaluate the extent of correctly expressed behavior, i.e., how many correct behaviors are expressed in the generated texts. We also use $F_1$ to evaluate behavioral performance. In this part, we use TAR (local) and trace (global) to reflect the model behaviors. As trace behaviors exist space explosion problem, thus, for trace F-measure, we only evaluate these models without loop behavior.

\begin{figure}[t]
  \centering
  \subfigure[$F_1$ score on TARs]{
    \label{sfig:tarF}
    \includegraphics[width=0.40\maxWidth]{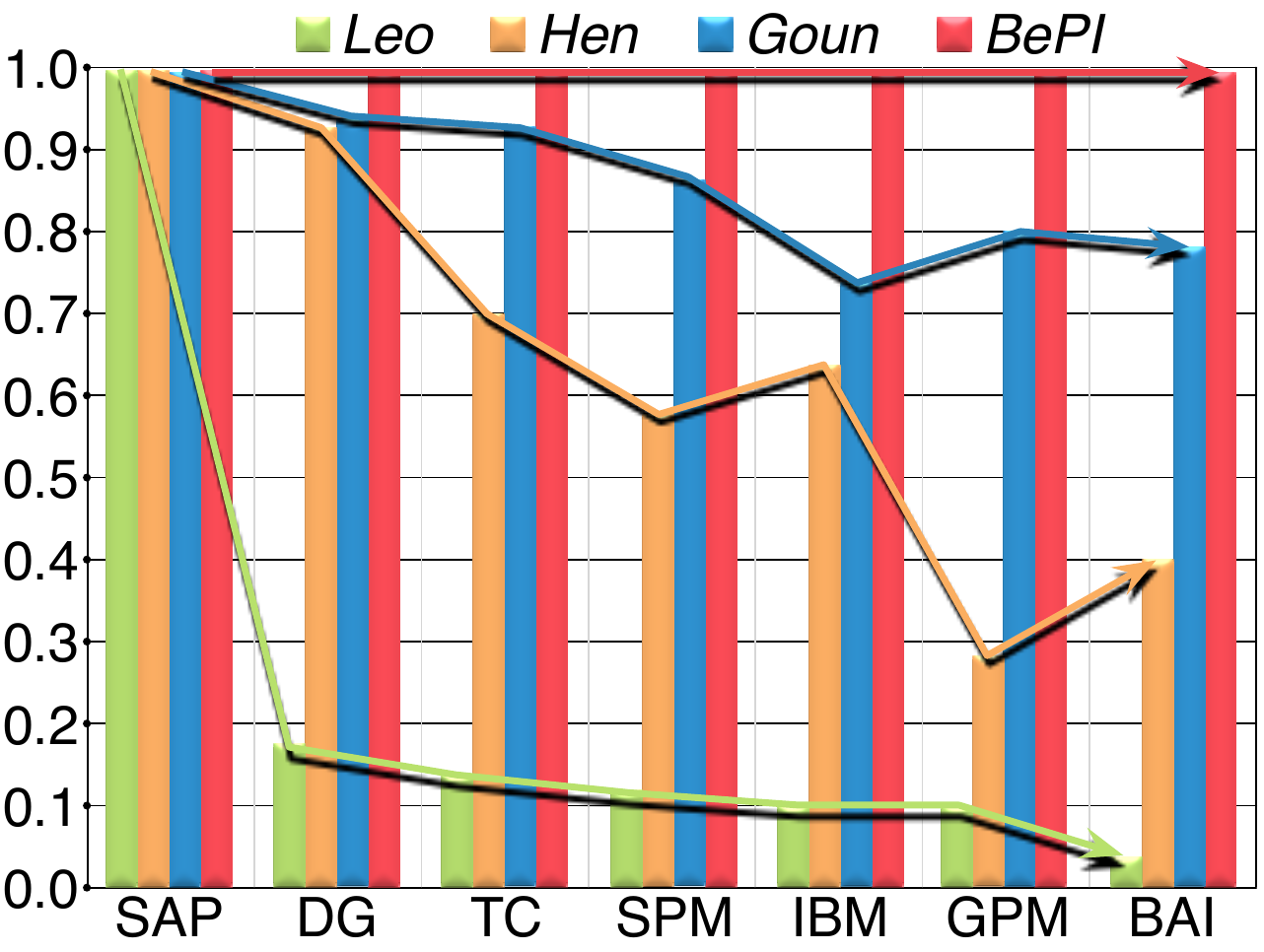}
  }
  \subfigure[$F_1$ score on traces]{
    \label{sfig:behaviorF}
    \includegraphics[width=0.40\maxWidth]{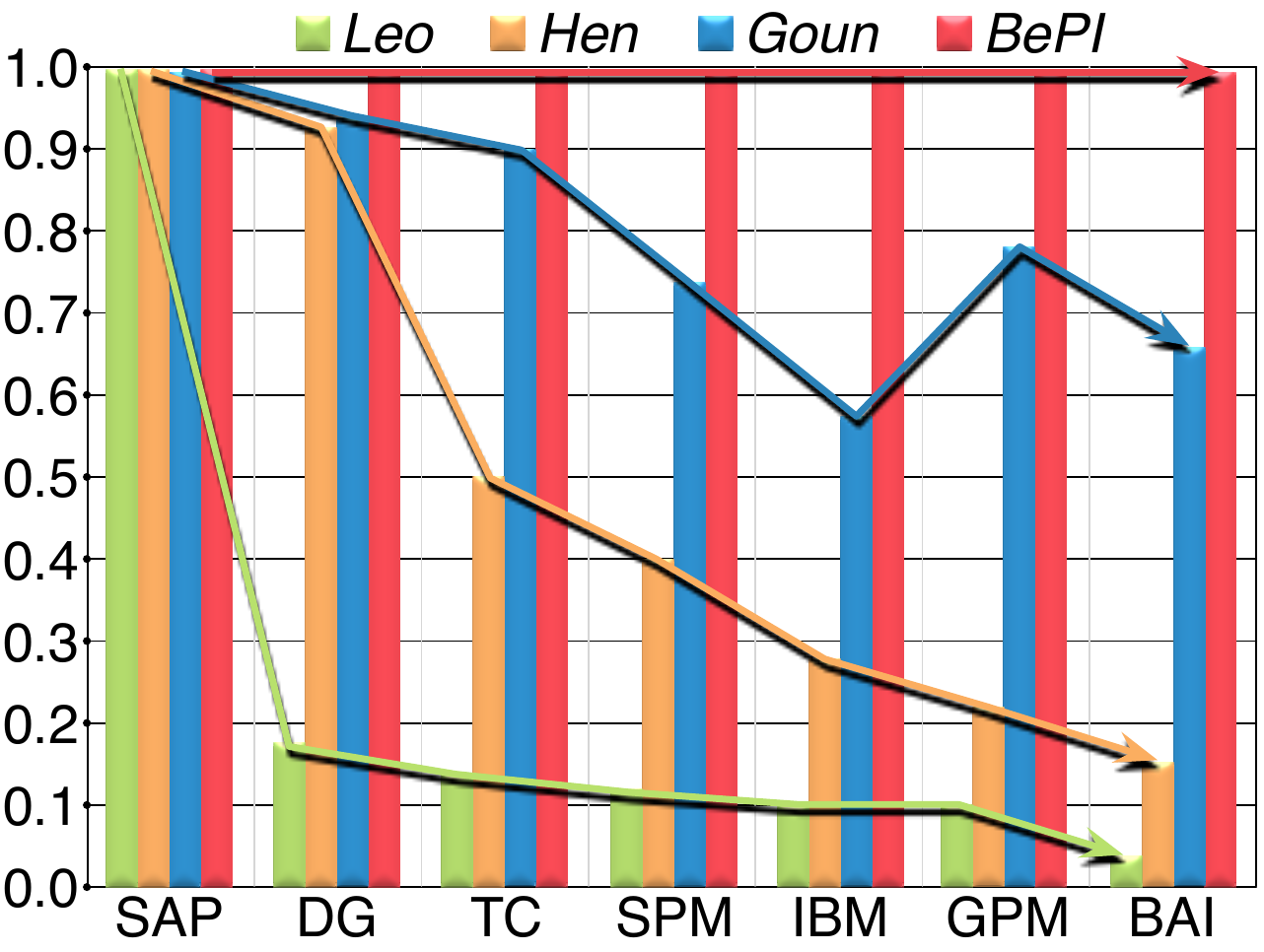}
  }
  \caption{The $F_1$ measures on behavioral dimensions.}
  \label{fig:behaviorMeasures}
  \end{figure}

Figure \ref{fig:behaviorMeasures} shows the results for the behavior dimensions (\emph{TAR, trace}). The results show that \emph{BePT} outperforms Leo, Hen and Goun significantly in terms of both TAR and trace performance. Leo performance falls sharply with decreasing $SMR$, while Hen and Goun drop more gently than Leo and they achieve around 70\% on BAI for trace $F_1$. \emph{BePT} gets the highest $F_1$ of around 100\% for both TAR and trace measures, and \emph{BePT} also produces a distinct improvement on TAR and trace $F_1$ over other methods. From these two performance results, we can conclude that \emph{BePT} showcases the best reproducibility over the state-of-the-art P2T methods.

\section{Conclusion and Future Work} \label{sec:conclusionAndDiscussion}

We present a behavior-based process translator. It first combines the structural and linguistic information into an RDT tree before decoding it by extracting the behavior paths. Then, we use NL tools to generate textual descriptions. Our experiments show the significant improvements that result on capability, detailedness, consistency, understandability and reproducibility. This approach can unlock the hidden value that lies in large process repositories in the cloud, and make them more reusable.

We also list some potential limitations of this study. Above all, when the model is unsound, \emph{BePT} informs the user that the model contains non-sound or wrong parts but without giving any correction advice. Another drawback concerns manual extraction of the NL text because of the limited number of participants. We cannot guarantee that each extraction rule for a generated text is identical. Thus, generating the correction advice and automatic reverse translation would also be of interest in future studies.

\section*{Appendices}
\begin{appendices}
  \section{The Proof of Behavior Correctness}\label{prf:correctness}
  \begin{property}
    Given a net system $S=(N,M)$ and its TAR set $\mathcal{T}(S)$. The behavior path set $\mathcal{P}$ of its CFP $\mathbb{N}$ by the linking rule (Definition \ref{def:linkingRule}) satisfies behavior correctness, $\forall \mathbb{P} \in \mathcal{P} \Rightarrow \mathcal{T}(\mathbb{P}) \subseteq \mathcal{T}(S)$.
    \end{property}
  
  \begin{qcProof}
    Given two Petri nets $N_i=(P_i, T_i, F_i), N_j=(P_j, T_j, F_j)$, we assume $\mathbb{P}=\wr \mathbb{S}_1, \mathbb{S}_2, \cdots, \mathbb{S}_n \wr$. Then, consider two situations: a) inside a single segment; b) between the linking of two segments:
  
    \begin{enumerate}[a)]
      \item The initial (default) marking $\bullet S$ is also the the initial marking of $\hbar(\mathbb{S}_1)$, i.e., the marking $\bullet \mathbb{S}_1$ is reachable. According to the definition of behavior segment, $\mathbb{S}_1\bullet$ is reachable from $\bullet\mathbb{S}_1$, and the firing rule guarantees $\mathcal{T}(\mathbb{S}_1) \subseteq \mathcal{T}(S)$. After executing $\mathbb{S}_1$, $\bullet \mathbb{S}_2$ is reachable as $\mathbb{S}_1 \bullet \supseteq \bullet \mathbb{S}_2$, so that $\mathcal{T}(\mathbb{S}_2) \subseteq \mathcal{T}(S)$ holds. Similarly, $\wr \mathbb{S}_1, \mathbb{S}_2 , \cdots, \mathbb{S}_{i-1} \wr \bullet \supseteq \bullet \mathbb{S}_i \Rightarrow \mathcal{T}(\mathbb{S}_{i}) \subseteq \mathcal{T}(S), i\in 1,2\cdots n$ holds.
  
      \item For two segments $\mathbb{S}_i, \mathbb{S}_{i+1}, i \in 1,2\cdots n-1$, we use the notation $\mathcal{T}(\mathbb{S}_i \wr \mathbb{S}_{i+1})$ to denote the TAR set in the joint points, i.e., $\mathcal{T}(\mathbb{S}_i \wr \mathbb{S}_{i+1}) = \{ a \prec b | a \in \bullet(\mathbb{S}_i\bullet) \wedge b \in (\bullet\mathbb{S}_{i+1})\bullet \}$. Since $\mathbb{S}_i \bullet \supseteq \bullet \mathbb{S}_{i+1}$ guarantees that $(\bullet\mathbb{S}_{i+1})\bullet$ can be fired after firing $\bullet(\mathbb{S}_{i}\bullet)$, i.e., $\mathcal{T}(\mathbb{S}_i \wr \mathbb{S}_{i+1}) \subseteq \mathcal{T}(S)$. Therefore, $\mathcal{T}(\wr \mathbb{S}_i, \mathbb{S}_{i+1} \wr) =\mathcal{T}( \mathbb{S}_i) \cup \mathcal{T}(\mathbb{S}_{i+1}) \cup \mathcal{T}(\mathbb{S}_i \wr \mathbb{S}_{i+1}) \subseteq \mathcal{T}(S), i\in 1,2\cdots n-1$.
      \end{enumerate}
  
    According to the above two points, we can conclude that $\forall \mathbb{P} \in \mathcal{P} \Rightarrow \mathcal{T}(\mathbb{P}) = \mathcal{T}(\mathbb{S}_1,\mathbb{S}_2,\cdots,\mathbb{S}_n) = \mathcal{T}(\mathbb{S}_1) \cup \mathcal{T}(\mathbb{S}_2) \cup \cdots \cup \mathcal{T}(\mathbb{S}_n) \cup \mathcal{T}(\mathbb{S}_1 \wr \mathbb{S}_2) \cup \mathcal{T}(\mathbb{S}_2 \wr \mathbb{S}_3) \cup \cdots \cup \mathcal{T}(\mathbb{S}_{n-1} \wr \mathbb{S}_n) \subseteq \mathcal{T}(S)$.
  \end{qcProof}

  \section{The Proof of Behavior Completeness}\label{prf:completeness}
  \begin{property}
    Given a net system $S=(N,M)$ and its TAR set $\mathcal{T}(S)$. The behavior path set $\mathcal{P}$ of its CFP $\mathbb{N}$ by linking rule (Definition \ref{def:linkingRule}) satisfies behavior completeness, $\forall \tau \in \mathcal{T}(S) \Rightarrow \exists \mathbb{P}\in\mathcal{P}, \tau \in \mathcal{T}(\mathbb{P})$.
  \end{property}
    
  \begin{qcProof}
    For any TAR $\tau= a \prec b \in \mathcal{T}(S)$, the place set $a \bullet \cap \bullet b$ is denoted as $\mathscr{P}$. The sub-model $( \mathscr{P}, \{a, b\}, \{a\} \times \mathscr{P} \cup \mathscr{P} \times \{b\} )$ is denoted as $\mathscr{N}$. We use $N_i \propto N_j$  to denote $P_i \subseteq P_j \wedge T_i \subseteq T_j \wedge F_i \subseteq F_j$, i.e., $N_i$ is a sub-model of $N_j$. Then, consider the following situations:
    
    \begin{enumerate}[a)]
      \item When $\forall p \in \mathscr{P}, p \notin \mathcal{SP}(\mathbb{N})$, there is no $p \in \mathscr{P}$ that can be the boundary node of a segment according to Definition \ref{def:segment} ($\mathcal{SP}$-bounded). Hence, $\mathscr{N}$ can only exist in the middle of a segment, i.e., $\exists \mathbb{S}_i, \mathbb{P}_j \Rightarrow \mathscr{N} \propto \mathbb{S}_i \propto \mathbb{P}_j \in \mathcal{P} \Rightarrow \tau \in \mathcal{T}(\mathscr{N}) \subseteq \mathcal{T}(\mathbb{P}_j)$.
    
      \item When $\forall p \in \mathscr{P}, p \in \mathcal{SP}(\mathbb{N})$, $\mathscr{P}$ is split, being the sink set of a certain segment $\mathbb{S}_i$ and the source set of a certain segment $\mathbb{S}_j$ ($\mathcal{SP}$-bounded), i.e., $\mathscr{P} = \mathbb{S}_i \bullet = \bullet \mathbb{S}_j$ always holds. Hence, $\exists \mathbb{S}_i, \mathbb{S}_j, \mathbb{P}_k \Rightarrow \mathscr{N} \propto \wr \mathbb{S}_i, \mathbb{S}_j \wr \propto \mathbb{P}_k \in \mathcal{P} \Rightarrow \tau \in \mathcal{T}(\mathscr{N}) \subseteq \mathcal{T}(\mathbb{P}_k)$.
    
      \item When $\exists p_1,p_2 \in \mathscr{P}, p_1 \notin \mathcal{SP}(\mathbb{N}), p_2 \in \mathcal{SP}(\mathbb{N})$, there is no $p \in \mathscr{P}$ can be the boundary node of a segment, or it contradicts Definition \ref{def:segment} (reply-hold). Hence, $\mathscr{N}$ can only exist in the middle of a segment, i.e., $\exists \mathbb{S}_i, \mathbb{P}_j \Rightarrow \mathscr{N} \propto \mathbb{S}_i \propto \mathbb{P}_j \in \mathcal{P} \Rightarrow \tau \in \mathcal{T}(\mathscr{N}) \subseteq \mathcal{T}(\mathbb{P}_j)$.
  
      \item When $\mathscr{P} = \varnothing$, i.e., $a$ and $b$ are in a concurrent relation. There always exists a concurrent split transition $t$. According to Definition \ref{def:segment}, $\exists \mathbb{S}_i \Rightarrow t \in \mathbb{S}_i \wedge a,b \in \mathbb{S}_i$ (reply-hold). Thus, $\exists \mathbb{S}_i, \mathbb{P}_j \Rightarrow \mathscr{N} \propto \mathbb{S}_i \propto \mathbb{P}_j \in \mathcal{P} \Rightarrow \tau \in \mathcal{T}(\mathscr{N}) \subseteq \mathcal{T}(\mathbb{P}_j)$.
    \end{enumerate}
  \end{qcProof}

  \section{The Proof of Description Minimality}\label{prf:ninimality}
  \begin{property}
    For a net system $S=(N,M)$, the pruned TARs $\mathcal{T}(S)$ by the Pruning Rule satisfies description minimality.
    \end{property}
  
  \begin{qcProof}
    According to Appendices \ref{prf:completeness}, for any TAR $\tau$, it can always be derived from a certain behavior path, i.e., $\forall \tau \in \mathcal{T}(S) \Rightarrow \exists \mathbb{P}\in\mathcal{P}, \tau \in \mathcal{T}(\mathbb{P})$. Hence, for two TARs $\tau_1, \tau_2$ of the original model with $\tau_1 \in \mathcal{T}(\mathbb{P}_i) \wedge \tau_2 \in \mathcal{T}(\mathbb{P}_j), i<j$. If $\tau_1\ne \tau_2$, $\{ \tau_1,\tau_2 \} \subseteq \mathcal{T}(S)$ always holds, while $\{ \tau_1 \} = \{ \tau_2 \} \subseteq \mathcal{T}(S)$ always holds if $\tau_1=\tau_2$. Therefore, the pruning rule always keeps TARs appearing at the first time, i.e., $\mathcal{T}(S)$ satisfies behavior minimality.
    \end{qcProof}
\end{appendices}

\section*{Acknowledgements}
The work was supported by the National Key Research and Development Program of China (No. 2016YFB1001101), the National Nature Science Foundation of China (No.71690231, No.61472207), and Tsinghua BNRist. We also would like to thank anonymous reviewers for their helpful comments.

\bibliographystyle{ACM-Reference-Format}
\bibliography{_BePT}

\end{document}